\documentclass[journal]{IEEEtran}

\ifCLASSINFOpdf
\else
  \usepackage[dvipdf]{graphicx}
\fi
 \usepackage{amssymb}
\usepackage{amsmath}
\usepackage{diagbox}
\usepackage{graphicx}
\usepackage{epstopdf}
\usepackage{booktabs}
\usepackage[justification=centering]{caption}
\usepackage{hyperref}

\usepackage{multirow}
\usepackage{slashbox}
\usepackage{color}
\usepackage{colortbl}
\usepackage{makecell}
\usepackage{algorithm}
\usepackage{algorithmic}

\usepackage{subfigure}
\usepackage{multicol}

 \renewcommand{\algorithmicrequire}{\textbf{Input:}} 
\renewcommand{\algorithmicensure}{\textbf{Output:}} 
\begin{document}
%


\title{An Effective Equivalence Model of Analyzing PLS of Multiple Eavesdroppers Facing Low-altitude Communication Systems}
%
%
%
\author{Yujia Zhao, Zhiyong Feng,~\IEEEmembership{Senior Member,~IEEE}, Kan Yu,~\IEEEmembership{Member,~IEEE}, Qixun Zhang,~\IEEEmembership{Member,~IEEE}, and Dong Li,~\IEEEmembership{Senior Member,~IEEE}
\thanks{This work is supported by the National Natural Science Foundation of China with Grant 62301076, the Macao Young Scholars Program with Grant AM2023015, and the Science and Technology Development Fund, Macau SAR, under Grant 0188/2023/RIA3.
}
\thanks{Y. Zhao, Z. Feng and Qixun Zhang are with the Key Laboratory of Universal Wireless Communications, Ministry of Education, Beijing University of Posts and Telecommunications, Beijing, 100876, P.R. China. E-mail: \{yjzhao0318@126.com, fengzy@bupt.edu.cn, zhangqixun@bupt.edu.cn\};}
\thanks{K. Yu (corresponding author) is with the Key Laboratory of Universal Wireless Communications, Ministry of Education, Beijing University of Posts and Telecommunications, Beijing, 100876, P.R. China; the School of Computer Science and Engineering, Macau University of Science and Technology, Taipa, Macau, 999078, P. R. China. E-mail: kanyu1108@126.com.}
\thanks{D. Li is with School of Computer Science and Engineering, Macau University of Science and Technology, Taipa, Macau, China. E-mail: dli@must.edu.mo.}


}

%
%

\markboth{IEEE Transactions on Vehicular Technology,~Vol.~, No.~, 2024}%
{Shell \Baogui Huang{\textit{et al.}}: Shortest Link Scheduling Under SINR}
%



\maketitle

\begin{abstract}
In low-altitude wireless communications, the increased complexity of wireless channels and the uncertainty of eavesdroppers (Eves)—caused by diverse altitudes, speeds, and obstacles—pose significant challenges to physical layer security (PLS) technologies based on fixed-position antennas (FPAs), particularly in terms of beamforming capabilities and spatial efficiency. In contrast, movable antennas (MAs) offer a flexible solution by enabling channel reconstruction through antenna movement, effectively compensating for the limitations of FPAs.
In this paper, we aim to derive a closed-form expression for the secrecy rate, a key metric in PLS, which is often unattainable in current studies due to the uncertainty of Eves. We construct an equivalent model that leverages the reconfigurable nature of MAs, equating the secrecy rates obtained by multiple Eves with single FPAs to those achieved by a single virtual Eve equipped with an MA array. To minimize the gap between these two types of secrecy rates, we formulate and solve an optimization problem by jointly designing \emph{the equivalent distance between the transmitter and the virtual Eve} and \emph{the antenna positions of MAs at the virtual Eve}.
Numerical simulations validate the effectiveness of the proposed equivalent model, offering a new perspective for PLS strategies. This work provides significant insights for network designers on how system parameters affect PLS performance.
\end{abstract}


%
\IEEEpeerreviewmaketitle

\section{Introduction}\label{intro}
The low-altitude economy is becoming a new growth point for China's future, particularly in critical applications such as logistics delivery, environmental monitoring, and emergency rescue \cite{JJSB202405070030}. The rapid development of the low-altitude economy relies on efficient and reliable wireless communications to ensure real-time information exchange among different control systems/devices. However, low-altitude wireless communications face significant security challenges due to two main reasons \cite{OJCS2024wei-UAV,TMC2024HUANG-LAC}: 1) wireless channels among devices are vulnerable to eavesdropping, tampering, and interference; 2) the devices used in the low-altitude economy are usually computing resource-constrained, making it difficult to implement traditional high-strength encryption technologies effectively. Therefore, exploring efficient and lightweight security solutions has become an urgent issue that needs to be addressed.

In this context, physical layer security (PLS), an emerging lightweight security technology, harnesses the randomness and uncertainty of wireless channels to enhance data transmission security. Unlike traditional encryption methods, PLS does not depend on computationally intensive algorithms, making it particularly well-suited for resource-constrained low-altitude devices. 
In the design of current PLS methods, which rely on fixed-position antenna (FPA) technology where all antennas are deployed at static locations, beamforming (BF) is used to enhance the received signal strength in desired directions while weakening it in undesired ones \cite{CM2024zhu-MA}. It is worth noting that PLS solutions developed for terrestrial communication networks are challenging to adapt to low-altitude wireless communication scenarios, due to the following reasons: 
\begin{itemize}
    \item \emph{Randomness and uncertainty of channel characteristics}: PLS relies on channel state information (CSI) to achieve the perfect secrecy. However, in low-altitude environments, factors such as altitude, speed, and obstacles can significantly impact the channel, which requires PLS solutions more adaptable to these fluctuations;
    \item \emph{Uncertainty of eavesdropper (Eve) location}: facing low-altitude environments, Eves may be positioned anywhere and could use a variety of devices to intercept communications.
\end{itemize}
To sum up, due to the strong directional feature of BF and mobility of the devices, low-altitude communications systems face pressing security challenges that need to be addressed imminently. For example, this strong directional BF of the confidential information is aimed at the Eves, there will be no security at all. To overcome this challenge, combing with the strong directional characteristic of the BF, exploiting BF as artificial noise (AN), which can be pointed to the Eves while minimizing the impact on the legitimate vehicles, is one of the most efficient methods. Subsequently, many other popular PLS methods are derived, such as joint design of: 1) BF-AN and UAV' trajectory \cite{WCL2019Li-BF-UAV}; 2) BF-AN and transmit power \cite{TWC2021li-AN-POWER}.

Due to the fixed geometry of FPA arrays, their steering vectors inherently exhibit correlation or orthogonality across different steering angles. This limits the flexibility of BF for interference nulling and beam coverage, making it challenging to simultaneously enhance signals in desired directions while mitigating interference in undesired ones. The ineffectiveness of FPAs in achieving this balance has been addressed by the authors in \cite{CL2024ma-MA-Multi-BF} and \cite{CL2023zhu-MA-Enhanced-BF}, showing that FPAs cannot amplify signals in desired directions and suppress interference in undesired directions at the same time \cite{arXiv2024dong-MA}.
Moreover, BF using FPAs fails to fully exploit the spatial degree of freedom (DoF), resulting in reduced security gains in both desired and undesired directions \cite{SPL2024hu-MA-PLS}. In light of these limitations, there is an urgent need for new and effective technologies to secure wireless communication in antenna based low-altitude systems.

Recently, movable antenna (MA) technology has been proposed and demonstrated the great potential to secure wireless communications by providing additional DoF \cite{ICASSP2024cheng-MA-PLS}. Unlike FPA, MAs offer the flexibility to dynamically adjust their orientation and configuration, and further realize more flexible BF. Accordingly, the optimal signal transmission and reception in dynamic environments can be achieved by varying the steering vectors corresponding to different angles, which can minimize the possibility of information leakage to the greatest extent possible, thereby enhancing the secrecy performance of low-altitude systems. Moreover, the potential Eves may move to the coverage blind zone of the conventional FPA arrays, at which they cannot guarantee the perfect secrecy by applying BF-AN methods, while MAs can flexibly adjust and compensate for the inherent defects of FPA to realize blind zone coverage\footnote{It is possible to apply MA to compensate FPA for coverage vulnerabilities, and a detailed theoretical analysis will be presented in the future work.}, making it more suitable for PLS in low-altitude systems. Therefore, MA empowered PLS design has great potential.

Inspired by the compelling benefits of MAs, there were two prior studies focusing on its use for PLS design, by jointing optimizing the BF vector and antenna positions of MAs at the base station (BS) for multiple colluding Eves \cite{SPL2024hu-MA-PLS}, and an Eve \cite{ICASSP2024cheng-MA-PLS}.
To reveal the fundamental secrecy rate limit of the MA-enabled or FPA-enabled secure communication system, they assumed that both the perfect CSI of all legitimate vehicles and Eves was available at the transmitter. However, the perfect CSI assumption remains a significant challenge in a low-altitude system because of the following reasons: 1) due to the mobility of the vehicles, the obtained CSI may be outdated; 2) Eves can adopt passive eavesdropping means to disguise the positions of themselves.
Considering an Eve and the effects of its imperfect CSI, in our another work \cite{arXiv2024feng-MA-PLS}, we established a framework of modeling it based on the concept of a virtual MA, under which the secrecy rate was maximized by jointly designing the BF and antenna positions of MAs at the transmitter. Numerical results  showed that the proposed PLS method can significantly enhance the secrecy rate compared to the FPA-based ones.

For the existence of multiple Eves with perfect CSI, it is still extremely challenging to establish the closed-form expression of the secrecy rate. Accordingly, the upper and lower bounds act as backup solutions \cite{TIFS2019yang-PLS-CSI,IOTJ2023illi-PLS-CSI}. More important, the closed-form expression of the secrecy rate can provide more significant insights into secure performance of low-altitude systems.
But it, as well as secure performance analysis, is still not clear. As a result, from the perspective of deriving closed-formed expression of the secrecy rate, \textbf{\emph{``is it feasible to approximate the secrecy rate achieved by multiple Eves with single FPAs as that of an Eve with a MA array?''}} need to be answered.

To address the aforementioned challenges and demonstrate the rationality
of the approximation, considering the scenario where a MA-enabled transmitter (Alice) transmits the confidentiality information to a legitimate receiver (Bob) with a single FPA, in the presence of multiple Eves with single FPA, based on the flexibility of MA arrays, we investigate equivalent model construction for two kinds of secrecy rates, which are achieved by multiple collusion Eves and an Eve with multiple MAs, respectively, and aim to minimize the equivalent deviation. That is, we approximate the channel quality of an Eve with an MA array, called as the virtual Eve, as that of multiple Eves with single FPA. Furthermore, the secrecy rate approximation minimization can be achieved by jointly optimizing the communication distance between the Alice and the virtual Eve and antenna positions of the MA deployed on the virtual Eve. The main contributions of this paper are summarized as follows.

\begin{itemize}
    \item First, to address the challenge of deriving the closed-form expression of the secrecy rate in the context of multiple Eves with single FPAs, we establish an equivalent model of the secrecy rate between multiple Eves and an Eve (denoted by virtual Eve) with a MA array, under which the different channel characteristics of multiple Eves can be effectively modeled by the mobility of the MAs;
    \item Next, with the help of alternative optimization, the secrecy rate deviation minimization is achieved by jointly optimizing the distance between the BS and virtual Eve and the antenna positions of MAs at the virtual Eve. In particular, the effectiveness of the proposed model can be validated by providing a lower bound of the secrecy rate achieved by multiple Eves with single FPAs, namely it can describe the worst performance of the system;
    \item Finally, the effectiveness and convergence of the proposed method are demonstrated by numerous results, in terms of the equivalent distance (i.e., the distance between the BS and virtual Eve) and equivalence difference minimization of secrecy rate. In particular, compared with individual optimization of the equivalent distance and antenna positions of MAs, the proposed method shows the best performance over different settings of the noise power, number of MAs and Eves, and path-loss exponent.
\end{itemize}

The rest of the paper is organized as follows. The related works are demonstrated in Section \ref{sec:related works}. In Section \ref{sec:network model}, the system model and performance metrics are presented, the optimization problem for secrecy rate deviation minimization is formulated. In Section \ref{sec:strategies}, the joint design of the equivalent distance and antenna positions of the MAs is utilized to achieve the secrecy rate deviation minimization, which is validated by simulated results in Section \ref{sec:evaluations}. Finally, conclusions and future works are discussed in Section \ref{sec:conclusions}.

\section{Related works}\label{sec:related works}

In the context of IoV systems, PLS methods are regarded as a critical and effective technology to ensure communication security between users. It’s noticed that most efforts in PLS schemes are based on FPA, and a lot of effective PLS schemes continuously appeared, such as the single/joint optimization of BF, transmission power optimization, and UAV trajectory \cite{IOTJ2022ruby-AN-POWER,JSTSP2019wang-BA-POWER} and \cite{ICC2022LYU-UAV-BF}.

\subsection{Fixed Position Antenna empowered PLS Methods}

Based on FPA technique, the AN becomes one of effective and popular methods in the study of the PLS. In particular, the AN signal can be utilized to point directly to the Eves based on the beamforming scheme, called as BF-AN in the following contents. The key idea of BF-AN is how to make a trade-off between the power for the data transmission and that of generating AN signal. 
Subsequently, in \cite{WCL2021qu-AN-POWER}, constrained by a fixed total power, Qu \emph{et al.} designed the optimal power allocation factor for above two kinds of powers, aiming at maximizing the secrecy rate. Similar works have done in \cite{IOTJ2022ruby-AN-POWER,JSTSP2019wang-BA-POWER}.
Whether the Eves' perfect CSI is known beforehand or not is an critical factor for ensuring the effectiveness of the PLS methods.
For the case that the perfect CSI of the Eves are known or not, in the coexistence of Bobs and Eves, in \cite{TVT2023chu-BF-AN-TP}, Chu \emph{et al.} designed two types of BF-AN PLS methods. When the Eves' perfect CSI is known, they jointly designed the transmitting beamformers of communication and radar systems to suppress the channel quality of Eves. On the other hand, when the Eves' CSI is unavailable, residual available power at the BS can be utilized to generate AN signals for disrupting the Eves while satisfying requirements of radar target detection. There are also similar works, such as BF design for Eves' perfect CSI \cite{TWC2024PALA-BF}, and or not \cite{TVT2021nan-BF-TP}.

Apart from the power allocation scheme for data signals and AN signals, combing BF-AN and UAV/RIS is another method to design effective PLS schemes, such as the joint design of UAV's trajectory and the beamformer of BF-AN for the secrecy rate maximization \cite{ICC2022LYU-UAV-BF,JSAC2021li-UAV-BF,TCOM2020wang-AN-UAV}.

However, due to immovable characteristics of the FPA, PLS methods based on BF-AN have two drawbacks as follows. 1) BF cannot be adjusted more accurately and flexibly; 2) spatial reuse and spatial diversity are relatively lower \cite{arXiv2024zeng-MA}.

\subsection{Movable Antenna empowered PLS Methods}
Compared to FPA, MA offers more higher spatial DoF and flexible BF design, which is regarded as a great potential technology for wireless PLS. Accordingly, several related works have been done \cite{SPL2024hu-MA-PLS,ICASSP2024cheng-MA-PLS,arXiv2024feng-MA-PLS,arXiv2024mei-MA-PLS,arXiv2024guan-MA-PLS}. In detail, the authors of \cite{SPL2024hu-MA-PLS} and \cite{ICASSP2024cheng-MA-PLS} addressed the problem of the secrecy rate maximization in the context of multiple Eves and a single Eve, respectively, by jointly optimizing the antenna position of the BSs' MA and beamformer. Different from previous methods of continuously optimizing antenna positions of the MA, for the security of an MA-enhanced multiple-input single-output (MISO) system, in \cite{arXiv2024mei-MA-PLS}, Mei \emph{et al.} constructed a discrete sampling point selection problem for antenna position optimization, and proposed a partial enumeration algorithm to solve it. Numerical results demonstrated that the proposed algorithm effectively improved the secrecy rate.
Considering the secrecy performance of the cell-free symbiotic radio systems, under which multiple distributed access points (APs) equipped with MAs collaboratively transmit confidential information to the primary user (PU), in the meanwhile the backscatter device (BD) transmits its own information to the
secondary user (SU) by reflecting incident signals from the APs, in \cite{arXiv2024guan-MA-PLS}, 
\emph{Guan et al.} aimed to maximize the secrecy rate of primary transmission by jointly optimizing the APs' transmission BF vectors and the antenna positions of the MAs deployed at the APs, while ensuring the quality of service for SUs.

In the previous works, to reveal the fundamental secrecy rate limit of the MA-enabled or FPA-enabled secure communication system, the location of the Eves usually was assumed to be
available and fixed at the transmitter \cite{TVT2022tang-PLS}, \cite{TVT2024ju-PLS}. However, the perfect assumption of known fixed-position Eve remains a significant challenge in an IoV system \cite{IOTJ2024chen-iov}. In addition, due to the vehicles movement, the obtained location estimation may be outdated \cite{arXiv2024feng-MA-PLS}. As a result, in our previous work \cite{arXiv2024feng-MA-PLS}, for the case of Eve's location uncertainty, we constructed an non-convex optimization problem for maximizing the secrecy rate, by jointly optimizing the BF of the BS and the antenna positions of the MAs, which can be further effectively addressed by utilizing simulated annealing and projected gradient ascent methods. Simulation results showed that the proposed method significantly improved the system's secrecy rate and achieved the perfect secrecy performance security.

\textbf{\emph{Notations:}} $\textbf{X}^T$ and $\textbf{X}^H$ represent the transpose and conjugate transpose of matrix $\textbf{X}$. \(x^*\) denotes the conjugate of the complex number \(x\). $ \mathbb{C}^{a \times b}$ denotes \(a \times b\) dimensional complex matrices. \(\rm{diag}(x)\) represents the diagonal matrix with diagonal elements  \(\mathbf{X}\). \(\left| \mathbf{x} \right|\) denotes taking the modulus of vector \(\mathbf{x}\). \([x]^{+}\) is  \(\max\{x, 0\}\). 
\(\mathbb{E}[x]\) represents expected operation for \(x\).

\section{System Model And Problem Formulation}\label{sec:network model}
In this section, we first present the system model for secure MA-assisted wireless communication. Then, Based on the concept of the MA array, we model the random location distributions of multiple Eves with a single MA via an virtual Eve with an MA array that is artificially placed at the system model and does not actually exist, corresponding significance and effectiveness are also given. Next, the equivalent model of the secrecy rate for multiple Eves with single FPA and the one obtained by an virtual Eve with an MA array is established. Finally, we construct an optimization problem that minimizes above equivalent model deviation in terms of the secrecy rate, by jointly optimizing the distance between the virtual Eve and the BS, and the antenna locations of MAs at the virtual Eve.

\subsection{MA-assisted System Model}\label{subsec:dis model}
\begin{figure}[htbp]
  \centering 
  \includegraphics[width=0.28\textwidth]{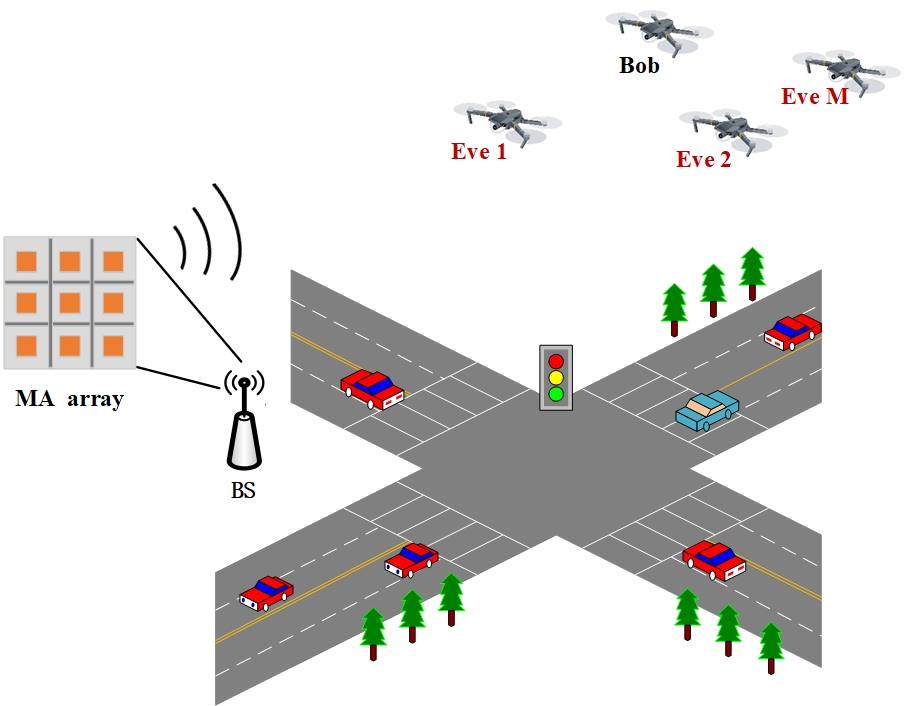}
  \caption{\small the system model with $M$ Eves} 
  \label{fig:system_model_multiple_Eves}
\end{figure}

We consider the uplink of an MA-assisted secure communications system, under which the BS equipped with $N_t>1$ transmitting MAs sends confidentiality information to a Bob with a single FPA, in the presence of $M$ Eves with a single FPA, as shown in Fig. \ref{fig:system_model_multiple_Eves}. 
The MAs are connected to RF chains via flexible cables, and can adjust their positions in a certain area \cite{CM2024zhu-MA}. But a minimum distance $d_{\min}$ between any two MAs is required to avoid the coupling effect between the antennas \cite{TWC2024ma-MA}. In addition, the MAs based on micro motors can achieve response speeds in level of \emph{ms} or even \emph{ns} \cite{arXiv2024zeng-MA}, then their switching delay can be ignored, when compared with the UAVs' speed\footnote{In particular, the rapid movement of the MAs may cause Doppler frequency shift. To solve the problem, the vehicles can reduce the corresponding effects via Doppler compensation technology.}.

\subsection{Equivalent Model of The Secrecy Rate between $M$ Collusion Eves and A virtual Eve with $M$ MAs, and Deviation Minimization}\label{subsec:equ model}
The locations of the Eves are randomly distributed, and they can collude with each other for achieving a better eavesdropping. Due to the location uncertainty of the Eves, the closed-form expression of the secrecy rate can hardly be derived, leading to less significant insights into secure wireless communications more effectively by adjusting key system parameters.
To solve this issue, we establish an equivalent model of the secrecy rate by approximating the performance of $M$ collusion Eves with single FPA via an Eve with $M$ MAs. In fact, such an Eve with an MA array does not exist, we call it as the virtual Eve. Then the approximation/equivalent model deviation is minimized by jointly designing the distance between the virtual Eve and the BS and the antenna locations of the MAs at the virtual Eve. Accordingly, we can conclude that the equivalent model deviation minimization can: 1) simplify the analysis of the secure performance in the context of multiple distributed Eves' collusion, thereby reducing the complexity of system modeling; 2) utilize MA technology to model the diverse channel characteristics of multiple Eves via adjusting antenna positions of the MAs at the virtual Eve.

\begin{figure}[htbp]
\centering  
\subfigure[\small the model of $M$ Eves with single FPAs]{
\includegraphics[width=0.35\textwidth]{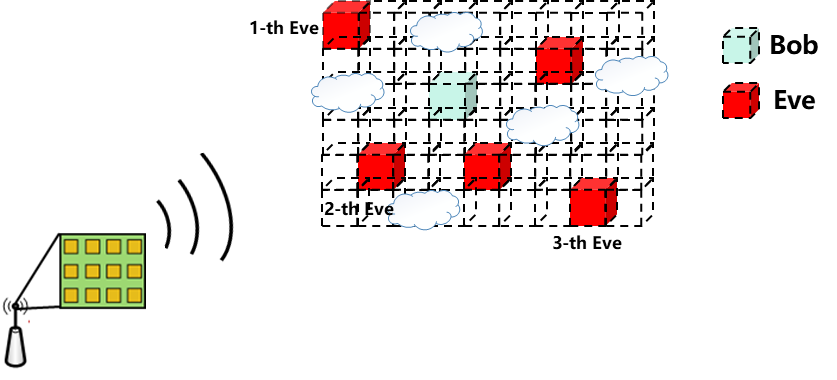}\hspace{5mm}\label{fig:sketch_1}}\\
\subfigure[\small the model of a virtual Eve with $M$ MAs]{
\includegraphics[width=0.35\textwidth]{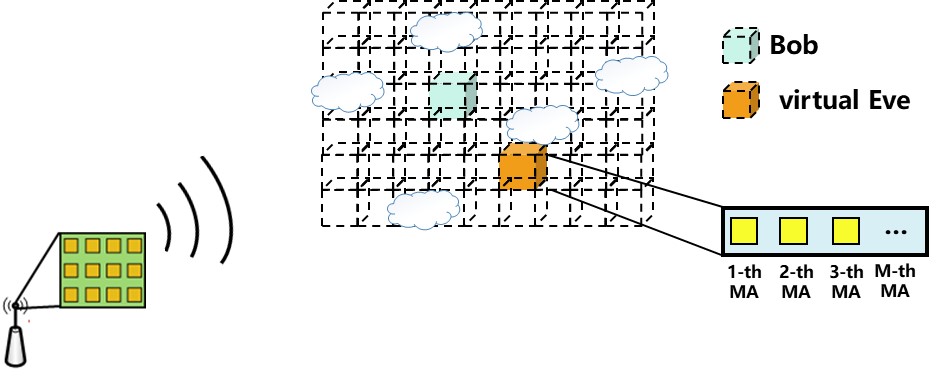}\label{fig:sketch_2}}
\caption{\small equivalent model between $M$ Eves and a virtual Eve with $M$ MAs}
\label{fig:equivalent model between two types of Eves} 
\end{figure}

As shown in Fig. \ref{fig:equivalent model between two types of Eves}, an equivalent model of $M$ Eves with single FPA via a virtual Eve with an MA array is presented. In this way, the BS with \(N\) transmitting MAs sends confidential information to the Bob, the virtual Eve equipping with $M$ MAs, instead of \(M\) Eves with single FPA, attempts to intercept this message. For the convenience of analysing, the MA array at the virtual Eve is deployed through a linear configuration, which means that the antenna only moves in one dimension.
Let $d_{\text{bob-bs}}$/$d_{\text{veve-bs}}$ be the communication distance between the BS and the Bob/virtual Eve.
The distance parameter $d_{\text{veve-bs}}$ and antenna locations of the MAs at the virtual Eve are crucial factors of the equivalent model deviation minimization, since they can affect the signal attenuation via diverse channel characteristics by changing the distance parameter and antenna positions of MAs \cite{TWC2024zhu-MA}, which provides an available method to model diverse channel features of multiple Eves with a single PFA via a virtual Eve with an MA array consisting of $M$ MAs.

By comparing the models in Fig. \ref{fig:sketch_1} and Fig. \ref{fig:sketch_2}, we aim to establish an equivalent model of the secrecy rates, which are obtained by $M$ collusion Eves with single FPA and a virtual Eve with $M$ MAs, respectively, and then minimize their deviation by jointly design of the distance between the BS and the virtual Eve and antenna positions of the MAs at the virtual Eve in the following analyses.

\subsection{The CSI of the Bob and Eves}
The perfect CSI is usually difficult to obtain. In this paper, we assume that the CSI of both the Bob and Eves is perfectly known, which can be justified as follows: 1) the Bob can transmit pilot signals to the BS, enabling precise channel estimation and allowing the BS to periodically update Bob's CSI; 2) While for the Eves, they may be legitimate users attempting to intercept other users' confidential information, thereby making it feasible to obtain Eve's CSI \cite{TVT2023chu-BF-AN-TP}.

\subsection{Channel Model}
A quasi-static flat-fading channel is considered between the BS equipped with an MA array and the Bob/Eves \cite{TWC2024ma-MA}. In detail, the MAs can move rapidly with the speed of light scale and the distance of wavelength scale. As a result, compared to the channel coherence time, the time that MAs take for the mobility can be also tolerable \cite{TWC2024ma-MA}. 
It is noteworthy that, different from the FPA, for MA-enabled secure communications, the channel can be reconstructed by adjusting the positions of MAs \cite{WCNC2024xiao-MA}. In other words, the channel vector is influenced not only by the propagation environment, but also affected by the positions of MAs. For convenience, based on the far-field assumption that the transmitter array size is much smaller than the signal propagation distance, we consider a field-response based channel model \cite{WCL2024li-MA}. As a result, the MAs at different positions have the same angle of arrival (AoA), angle of departure (AoD), and complex amplitude coefficients, while only the phase of different channel paths varies with the MAs' positions.

\begin{figure}[htbp]
  \centering \includegraphics[width=0.25\textwidth]{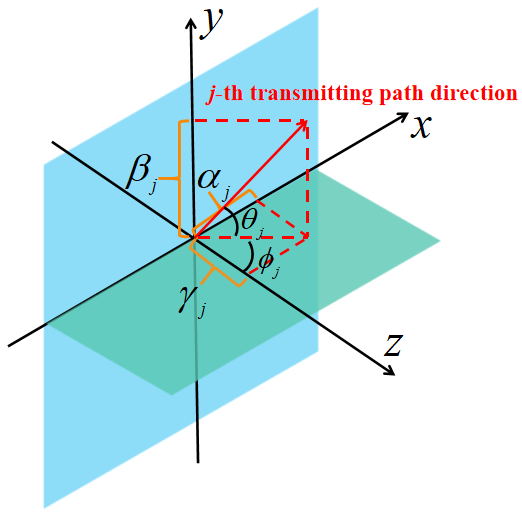}
  \caption{\small spatial angles of the $j$-th transmitting path} 
    \label{fig:Spatial angles}
\end{figure}

First, we establish the transmission field response vector/matrix between the BS with $N$ MAs and the Bob as follows.
Let \( \mathbf{t}_n = [x_n, y_n, z_n] \) and \( \mathbf{T} = [\mathbf{t}_1, \mathbf{t}_2, \dots, \mathbf{t}_N] \) be the position of the \( n \)-th transmitting MA at the BS, and the position matrix of the \( N \) MAs at the BS, respectively. Due to the existence of multiple paths between MAs and Bob, we assume that there are \( L^t \) transmission paths between the MAs and the Bob. Accordingly, we can utilize the vertical angle \( \theta \) and the horizontal angle \( \phi \) to describe the direction of the \( j \)-th path. In this way, the vertical and horizontal AoD of the \( j \)-th path of the MA are \( \theta_j \) and \( \phi_j \) (\( 1 \leq j \leq L^t \)), respectively. As shown in Fig. \ref{fig:Spatial angles}, given a three-dimensional coordinate system, the virtual AoD can be defined as the projection of the normalized direction vector in the three-dimensional coordinate system \cite{TWC2024ma-MA}. Thus, the normalized direction vector of the \( j \)-th path in the three-dimensional coordinates can be represented as \( \mathbf{p}^j = [\cos\theta_j\cos\phi_j, \cos\theta_j\sin\phi_j, \sin\theta_j] \), and the vertical and horizontal AoD angles (i.e., \( \theta_j \) and \( \phi_j \)) vary in the range of \( [-\pi/2, \pi/2] \). In addition, based on the three-dimensional vector projection, the MA can be represented as moving in the xy-plane. Let \([0,0,0]\) be the origin of the $n$-th MA. When it to a new position, the corresponding position vector can be denoted as \( \mathbf{t}_n = [x_n, y_n, 0] \). In this way, the distance difference and phase difference of the \( j \)-th path are \(  {\mathbf{t}_n}^T \mathbf{p}^j\) and \(\frac{2\pi}{\lambda}{\mathbf{t}_n}^T \mathbf{p}^j \), where \( \lambda \) represents the wavelength.

Therefore, the received transmission field response vector of the $n$-th transmitting MA at the Bob can be expressed as
\begin{displaymath}
\mathbf{g}_n (\mathbf{t}_n) = \left[ e^{j\frac{ 2\pi}{\lambda} \mathbf{t}_n^{T} \mathbf{p}^1}, e^{j\frac{ 2\pi}{\lambda} \mathbf{t}_n^{T} \mathbf{p}^2}, \ldots, e^{j\frac{ 2\pi}{\lambda} \mathbf{t}_n^{T} \mathbf{p}^{L^t}} 
\right]^T, 1 \leq n \leq N .
\end{displaymath}
The corresponding field response matrix can be represented as
\begin{displaymath}
\mathbf{G}(\mathbf{T}) = \left[ \mathbf{g}_1(\mathbf{t}_1), \mathbf{g}_2(\mathbf{t}_2), \ldots, \mathbf{g}_N(\mathbf{t}_N) \right] \in \mathbb{C}^{L^t \times N}.
\end{displaymath}

Next, we model the eavesdropping channels of $M$ Eves and virtual Eve as follows.

\emph{The channel model of $M$ randomly distributed Eves}: Based on the field response channel model, for $M$ Eves, the transmission field response matrix at the BS remains is \( \mathbf{G(T)} \). Due to the FPA deployed at the Eves, the field response vector at any Eves can be uniformly represented as \( \mathbf{f} = [1,1,\ldots,1]^T \in \mathbb{C}^{L_t \times 1} \) \cite{ICASSP2024cheng-MA-PLS}. Furthermore, we assume that the number of transmission paths and reception paths from the BS to every Eve is the same, namely \( L^{r}_{\rm eve} = L^t\), since the geometry channel model. Let \( \mathbf{\Sigma}_m = \text{diag}(\sigma_{1,m}, \sigma_{2,m}, \ldots, \sigma_{L^t,m}) \in \mathbb{C}^{L^{r}_{\rm eve} \times L^t} \) be the path response matrix of the \( m \)-th Eve, where \( \sigma_{l,m} \) represents the complex gain of the \( l \)-th path. 

Therefore, according to the field response channel model, the field response channel matrix from the BS to the \( m \)-th Eve is given by
\begin{equation}
\begin{aligned}
\mathbf{h}_m^e(\mathbf{T}) &= \left( \mathbf{f}^T \mathbf{\Sigma}_m \mathbf{G}_m(\mathbf{T}) \right)^T \\
&= \left[ h_m^e(\mathbf{t}_1), h_m^e(\mathbf{t}_2), \ldots, h_m^e(\mathbf{t}_N) \right]^T \in \mathbb{C}^{N \times 1},
\label{eq:h(T)}
\end{aligned}
\end{equation}
where \(h_m^e(\mathbf{t}) = \sum\limits_{l=1}^{L^t} \sigma_l e^{\frac{j 2\pi}{\lambda} \mathbf{t}^T \mathbf{p}_m^l} \) is the sum of weights for the field response vectors of all paths caused by different antenna positions of the MAs \cite{ICASSP2024cheng-MA-PLS}. To sum up, the signal received at the \( m \)-th Eve can be expressed as
\begin{equation*}
y_m^e = \mathbf{h}_m^e(\mathbf{T})^H \mathbf{w} x + n_e, \quad 1 \leq m \leq M,
\end{equation*}
where \( \mathbf{w}x \) represents the data transmitted by the BS, \( x \) denotes the data symbol, \( \mathbf{w} \) represents the BF vector designed at the BS for transmitting information to the Bob, and \( n_e \) is the noise received at the $m$-th Eve, which follows a complex Gaussian distribution with the mean of 0 and noise power of \( \sigma^2 \). Accordingly, the Signal-Noise Ratio (SNR) at the \( m \)-th Eve can be expressed as
\begin{equation*}
\gamma_m^e = \frac{\left| \mathbf{h}_m^e(\mathbf{T})^H \mathbf{w} \right|^2}{\sigma^2}.
\end{equation*}
Furthermore, the channel capacity of all \( M \) Eves with eavesdropping collision can be represented as 
\begin{equation}\label{eq:channel capacity eves}
C_{\text{col}}^e = \log_2 ( 1 + \sum_{m=1}^{M} \gamma_m^e ).
\end{equation}

\emph{The channel model of the virtual Eve:} Similarly, the transmission field response matrix at the BS remains is \( \mathbf{G(T)} \). Let \( r_m=[x_m] \) denote the x-axis position coordinate of the \( m \)-th antenna of a linear MA array at the virtual Eve. The position matrix of $M$ MAs at the linear array can be denoted as \( \mathbf{R} = [r_1, r_2, \dots ,r_M] \). The field response vector received at the position \( r_m \) can be expressed as

\begin{equation*}
\mathbf{f}(r_m) = \left[ e^{j\frac{ 2\pi}{\lambda} r_m \cos \theta_1}, e^{j\frac{ 2\pi}{\lambda} r_m \cos \theta_2}, 
\ldots, e^{j\frac{ 2\pi}{\lambda} r_m \cos \theta_{L^r}} 
\right]^T,
\end{equation*}
where \( L^r \) represents the number of receiving paths for VEve, and \( \theta_i \) denotes the AoA for the \( i \)-th path of the equivalent Eve, with values ranging within \( [0, \pi] \).

Next, the field response channel matrix of the \( M \) receiving antennas at the virtual Eve can be expressed as
\begin{equation*}
\mathbf{F}(\mathbf{R}) = \left[ \mathbf{f}(r_1), \mathbf{f}(r_2), \ldots, \mathbf{f}(r_M) \right] \in \mathbb{C}^{L^t \times M}.
\end{equation*}
Similarly, the number of transmission paths and reception paths from the BS to the virtual Eve is the same, namely \( L^{r}_{\rm veve} = L^t\). The corresponding path response matrix is \( \mathbf{\Sigma}^e = \text{diag}(\sigma_{1}, \sigma_{2}, \ldots, \sigma_{L^t}) \in \mathbb{C}^{L^{r}_{\rm veve} \times L^t} \). As a result, the channel matrix at the virtual Eve is \( \mathbf{h}^e(\mathbf{T}, \mathbf{R}) = \left( \mathbf{F}(\mathbf{R})^H \mathbf{\Sigma}^e \mathbf{G}(\mathbf{T}) \right)^T \), and the received signal can be expressed as
\begin{equation*}
y^e = \mathbf{h}^e(\mathbf{T}, \mathbf{R})^H \mathbf{w} x + n_e.
\end{equation*}
Accordingly, the corresponding SNR at the virtual Eve can be expressed as
\begin{equation*}
\gamma^{ve} = \frac{\left| \mathbf{h}^e(\mathbf{T}, \mathbf{R})^H \mathbf{w} \right|^2}{\sigma^2}.
\end{equation*}
Furthermore, the channel capacity of the virtual Eve with a linear MA array can be expressed as
\begin{equation*}
C_{\text{MA}}^{ve} = \log_2( 1 + \gamma^{ve}).
\end{equation*}

\subsection{Performance Metric: Equivalence Difference Minimization of Secrecy Rate}\label{sub:equivalence difference minimization}
In this paper, we answer the question of \textbf{\emph{``is it feasible to approximate the secrecy rate achieved by multiple Eves with a single FPA as that of an Eve with a MA array?''}}. Therefore, we aim to minimize the difference of the secrecy rate obtained by $M$ Eves with single FPA and a virtual Eve with $M$ MAs. Let $C_{\rm bob}$ be the channel capacity of the Bob. The secrecy rate difference after using the equivalent model in Subsection \ref{subsec:equ model} can be represented as 
\begin{equation}
\begin{aligned}
\Delta R_{\text{sec}}& = \left[ C_{\text{bob}} - C^{ve}_{\text{MA}} \right]^{+} - \left[ C_{\text{bob}} - C^e_{\text{col}} \right]^{+}\\
&\stackrel{(a)}{=} C^{ve}_{\text{MA}} - C^e_{\text{col}}.
\label{equ:delta_sec}
\end{aligned}
\end{equation}
In order to ensure the perfect secrecy of the system after the equivalent model being established, i.e., approximating the secrecy rate obtained by the virtual Eve with $M$ MAs as that of $M$ collusion Eves, the equivalent model should consider the worse case where the virtual Eve has a better channel quality than that of all $M$ collusion Eves. As a result, the secrecy rate obtained by the virtual Eve with an MA array should be lower than that of $M$ Eves with single FPA. Therefore, $(a)$ holds. Moreover, to ensure the effectiveness of the equivalent model in Subsection \ref{subsec:equ model}, we aim to make the secrecy rate obtained by the virtual Eve as close as possible to that of $M$ collusion Eves. That is, \textbf{$\Delta R_{\text{sec}}\geq 0$ and need be minimized}.

\subsection{Problem Formulation}

To simplify the problem analysis, form the expression of $C^{ve}_{\text{MA}}$ and $C^e_{\text{col}}$, the minimization of $\Delta R_{\rm sec}$ is equivalent to the difference minimization of SNRs at the virtual Eve and $M$ collusion Eves. Therefore, we can get
\begin{equation}
\Delta \text{SNR}_{\rm dif} = \text{SNR}_{\rm veve}- \text{SNR}_{\text{col}}.
\label{equ:delta_SNR}
\end{equation}
where $\rm{SNR}_{\rm veve}=\it\gamma^{ve}$ and $\rm{SNR}_{\text{col}}=\it\sum_{m=\rm{1}}^{M} \gamma_m^e$.  
For the channel matrix in the expressions of Eq. \eqref{eq:h(T)} and Eq. \eqref{eq:H(T,R)} 
 (on the top of the next page), the \(\sigma_l\) and \(\sigma_u\) are random variables from the path response matrix. $\rm{SNR}_{\rm veve}$ and $\rm{SNR}_{\text{col}}$ are functions of the r.v. \(\sigma_l\) and \(\sigma_u\), leading to $\Delta \rm SNR$ being a random variable. For the reason, we turn to solving \textbf{the minimization of $\Delta \rm SNR_{dif}$' expectation}. From the previous analysis, the distance \( d \) between the BS and the virtual Eve, and the antenna position matrix \( \mathbf{R} \) of the linear MA array at the virtual Eve are critical factors influencing the effectiveness and correctness of the equivalent model. Therefore, the joint design of \( d \) and \( \mathbf{R} \) for $\Delta \rm SNR$' expectation can be formulated as follows.
\begin{subequations}\label{eq:optimal inital}
\begin{align}
\text{Objective:}~~& \min_{d, \mathbf{R}} \mathbb{E}[\Delta \text{SNR}_\text{dif}]  \label{objecive init}\\
\text{s.t. } &  \mathbf{r}_n \in \Psi_n \quad \text{for} \quad n = 1, 2, \ldots, N  \label{theparentequation b},\\
&  \left| \mathbf{r}_n - \mathbf{r}_{n'} \right| \geq D_{\min}  \quad \text{for} \quad n \neq n'\label{theparentequation c},\\
&  \mathbb{E}[\Delta \text{SNR}_\text{dif}] \geq 0,\label{theparentequation e}
\end{align}
\end{subequations}
where constraint \eqref{theparentequation b} represents the allowable movement range of the MAs at the virtual Eve, constraint \eqref{theparentequation c} represents the minimum distance between two antennas, where \( D_{\min} \) denotes the minimum safety distance between any two MAs for avoiding the coupling effect.
Note that Eq. \eqref{objecive init} is non-convex optimization problem, due to the following reasons: 1) the non-convexity of objective function with respect to $(d,\textbf{R})$; 2) the non-convex minimum distance constraint of \eqref{theparentequation c} and non negative constraint conditions \eqref{theparentequation e} for the objective function. Moreover, the equivalent distance \(d\) is coupled with VMA antenna position \(\mathbf{R}\), which makes the problem more challenging to be solved. To the best of the authors' knowledge, this is the first work that takes into account the equivalence difference for PLS empowered by the MA technology. Although there is no general approaches (e.g., taylor expansion etc.) to solving problem Eq. \eqref{objecive init} optimally, in the following section, we propose an effective algorithm to find a globally optimal solution for problem Eq. \eqref{objecive init}.

\textcolor{blue}{\begin{figure*}
\label{equ:H(T,R)w}
\begin{align}
{\left| \mathbf{h}^{e}(\mathbf{T}, \mathbf{R})^{H} \mathbf{w} \right|^{2}} = 
\left|\left[
\begin{array}{ccc}
w_1\sum\limits_{u=1}^{L^t} \sigma_u e^{j \frac{2\pi}{\lambda} (-r_1 \cos \theta_u + t_1 \mathbf{p}^u)} + \cdots + w_N\sum\limits_{u=1}^{L^t} \sigma_u e^{j \frac{2\pi}{\lambda} (-r_1 \cos \theta_u + t_N \mathbf{p}^u)} \\
\vdots \\
w_1\sum\limits_{u=1}^{L^t} \sigma_u e^{j \frac{2\pi}{\lambda} (-r_M \cos \theta_u + t_1 \mathbf{p}^u)} + \cdots + w_N\sum\limits_{u=1}^{L^t} \sigma_u e^{j \frac{2\pi}{\lambda} (-r_M \cos \theta_u + t_N \mathbf{p}^u)}
\end{array}
\right] \right|^{2}.
\end{align}
\hrule 
\end{figure*}}
\subsection{Optimization Problem Simplification}
As described above, the expectation operation in the objective function Eq. \eqref{objecive init} makes the optimization problem more difficult to solve. In this subsection, we derive a general expression of the objective function by utilizing the linearity of expectation. Then, the objective function can be decomposed into two terms as follows: 
\begin{equation}
\mathbb{E}[\Delta \text{SNR}_\text{dif}]= \mathbb{E} \left[ \text{SNR}_{\text{veve}}\right] - \mathbb{E} \left[ \text{SNR}_{\text{col}} \right].
 \label{equ:vma-dis}
\end{equation}
Next, we first derive the expected value of the SNR received at the virtual Eve. On the one hand, the corresponding channel matrix, i.e., \(\mathbf{h}^{e}(\mathbf{T}, \mathbf{R})^{H} \), can be expression shown on the top of the next page.
\begin{figure*}
\begin{align}
\mathbf{h}^e(\mathbf{T}, \mathbf{R})^H = 
\left[
\begin{array}{ccc}
\sum\limits_{u=1}^{L^t} \sigma_u e^{j \frac{2\pi}{\lambda} (-r_1 \cos \theta_u + t_1 \mathbf{p}^u)} & \cdots & \sum\limits_{u=1}^{L^t} \sigma_u e^{j \frac{2\pi}{\lambda} (-r_1 \cos \theta_u + t_N \mathbf{p}^u)} \\
\vdots & \ddots & \vdots \\
\sum\limits_{u=1}^{L^t} \sigma_u e^{j \frac{2\pi}{\lambda} (-r_M \cos \theta_u + t_1 \mathbf{p}^u)} & \cdots & \sum\limits_{u=1}^{L^t} \sigma_u e^{j \frac{2\pi}{\lambda} (-r_M \cos \theta_u + t_N \mathbf{p}^u)}
\end{array}
\right] \in \mathbb{C}^{M \times N}.\label{eq:H(T,R)}
\end{align}
\hrule 
\end{figure*}
Let \(\mathbf{w} = \left[ w_1, \ldots, w_n, \ldots, w_N \right]^T \in \mathbb{R}^{N \times 1} \) denote the beamforming vector for the Bob, where \(w_n\) represents the beamforming components of the \( n \)-th transmitting antenna. In this way, the power strength of eavesdropping signal, i.e., \( {\left| \mathbf{h}^{e}(\mathbf{T}, \mathbf{R})^{H} \mathbf{w} \right|^{2}} \), can be expressed as shown on the top of the next page.

On the other hand, to simplify the notation, let
\(\alpha_u(r_m) = e^{-j \frac{2\pi}{\lambda} (-r_m \cos \theta_u)} \), and \(\beta_u(\mathbf{t}_n) =  e^{-j \frac{2\pi}{\lambda} (\mathbf{t}_n \mathbf{p}^u)}\). Then, we can get \(\gamma_u = w_1 \beta_u(\mathbf{t}_1) + \ldots + w_N \beta_u(\mathbf{t}_N), 1 \leq u \leq L^t\). Also, the power strength of eavesdropping signal, namely ${\left| \mathbf{h}^{e}(\mathbf{T}, \mathbf{R})^{H} \mathbf{w} \right|^{2}}$, can be rewritten as: 
\begin{displaymath}
\begin{aligned}
 &{\left| \mathbf{h}^{e}(\mathbf{T}, \mathbf{R})^{H} \mathbf{w} \right|^{2}}=
\left|\left[
\begin{array}{c}
\sum\limits_{u=1}^{L^t} \sigma_u \alpha_u(r_1) \gamma_u \\
\vdots \\
\sum\limits_{u=1}^{L^t} \sigma_u \alpha_u(r_M) \gamma_u
\end{array}
\right]\right|^2\\
&=
\left|\left( \sum\limits_{u=1}^{L^t} \sigma_u \alpha_u(r_1) \gamma_u \right)\right|^2 + \ldots + \left|\left( \sum\limits_{u=1}^{L^t} \sigma_u \alpha_u(r_M) \gamma_u \right)\right|^2.
\end{aligned}
\end{displaymath}
Thus, the expectation of the SNR \( \mathbb{E} \left[ \rm SNR_{\text{veve}}\right] \) measure by the virtual Eve with $M$ MAs can be represented as
\begin{displaymath}\label{equ:EE}
\mathbb{E} \left[ \frac{
\left|\left( \sum\limits_{u=1}^{L^t} \sigma_u \alpha_u(r_1) \gamma_u \right)\right|^2 + \ldots + 
\left|\left( \sum\limits_{u=1}^{L^t} \sigma_u \alpha_u(r_M) \gamma_u \right)\right|^2
}{\sigma^2} \right],
\end{displaymath}
which can be simplified into the following equation by using the basic properties of expectation:
\begin{equation}
\mathbb{E}[\text{SNR}_{\text{veve}}] = \frac{1}{\sigma^2} \left\{ \sum_{z=1}^{M} \mathbb{E} \left[ \left|\left( \sum_{u=1}^{L^t} \sigma_u \alpha_u(r_z) \gamma_u \right)\right|^2 \right] \right\}.
\label{equ:SNRvma}
\end{equation}
The method for solving each expectation term in Eq. \eqref{equ:SNRvma} is similar. Taking the first term as an example (for \( z = 1 \)), we deal with the expression of \( E_1 = \mathbb{E}\left[\left|\left( \sum\limits_{u=1}^{L^t} \sigma_u \alpha_u(r_1) \gamma_u \right)\right|^2\right] \). Based on the definition of the modulus squared of a complex number, we can have
\begin{equation}
\begin{aligned}
E_1 &= \mathbb{E}\left[\left( \sum_{u=1}^{L^t} \sigma_u \alpha_u(r_1) \gamma_u \right) \left( \sum_{v=1}^{L^t} \sigma_v \alpha_v(r_1) \gamma_v \right)^*\right]\\
&= \mathbb{E} \left[ \sum_{u=1}^{L^t} \sum_{v=1}^{L^t} \sigma_u \alpha_u(r_1) \gamma_u \sigma_v^* \alpha_v^*(r_1) \gamma_v^* \right]\\
&= E_{u=v} + E_{u \neq v},
\label{equ:E1}
\end{aligned}
\end{equation}
where the parameter \(\sigma\) in the path response matrix is independent and identically distributed (i.i.d.), which follows a complex Gaussian distribution with the mean of zero and variance \(\frac{g_0 d^{-\alpha}}L\), \(g_0\) denotes the average channel gain at a reference distance of 1 meter, \(L\) is the number of transmitting/receiving paths, \(d\) represents the distance between the BS and the Bob/virtual Eve, and \(\alpha\) is the path-loss exponent \cite{TWC2024zhu-MA-}. Moreover, \(E_{u=v} = \mathbb{E} \left[ \sum\limits_{u=v=1}^{L^t} \sigma_u \alpha_u(r_1) \gamma_u \sigma_u^* \alpha_u^*(r_1) \gamma_u^* \right]\) and \(E_{u\neq v} = \mathbb{E} \left[ \sum\limits_{u \neq  v} \sigma_u \alpha_u(r_1) \gamma_u \sigma_v^* \alpha_v^*(r_1) \gamma_v^* \right]\).
Therefore, for the second term in Eq. \eqref{equ:E1}, we can conclude that \(E_{u \ne v} = 0\), since the parameter \(\sigma\) is corresponding to different paths and is independent with the mean of zero. Next, we solve for the expression of \(E_{u = v}\) as follows:
\begin{equation}
\begin{aligned}
E_{u=v} = \mathbb{E} \left[ \left| \sigma_u \right|^2 \right] \left| \sum_{u=1}^{L^t} \alpha_u(r_1) \gamma_u \right|^2 
= \frac{g_0 d^{-\alpha}}{L^t} \left| \sum_{u=1}^{L^t} \alpha_u(r_1) \gamma_u \right|^2,
\label{equ:Eu=v}
\end{aligned}
\end{equation}
To sum up, Eq. \eqref{equ:E1} is derived for $z=1$. By applying the same method to deal with the expressions for \(z=2, \ldots, M\) in Eq. \eqref{equ:SNRvma}, then the expression of \(\mathbb{E}[\text{SNR}_{\text{veve}}]\) can be represented as
\begin{equation}
\mathbb{E}[\text{SNR}_{\text{veve}}] = \frac{g_0 d^{-\alpha}}{L^t \sigma^2} \sum_{z=1}^{M} \left| \sum_{u=1}^{L^t} \alpha_u(r_z) \gamma_u \right|^2.
\end{equation}

Furthermore, we solve the expected value \(\mathbb{E}[\text{SNR}_{\text{col}}]\) in Eq. \eqref{equ:vma-dis}, which represents the SNR of $M$ Eves with single FPA. The power strength of received signal is given by
\begin{equation*}
\begin{aligned}
\left| \mathbf{h}_m^e(\mathbf{T})^H \mathbf{w} \right|^2 
&= \left| w_1\sum_{u=1}^{L^t}   \sigma_u \beta_u(\mathbf{t}_1) + \ldots + w_N \sum_{u=1}^{L^t}\sigma_u \beta_u(\mathbf{t}_N)  \right|^2 \\
&= \left| \sum_{u=1}^{L^t} \sigma_u \gamma_u \right|^2.
\end{aligned}
\end{equation*}
By using the same method for solving \(\mathbb{E}[\text{SNR}_{\text{veve}}]\), the expression of \(\mathbb{E}[\text{SNR}_{\text{col}}]\) can be described as
\begin{equation}
\mathbb{E}[\text{SNR}_{\text{col}}] = \sum_{m=1}^{M} \frac{g_0 d_m^{-\alpha}}{L^t \sigma^2} \left( \sum_{u=1}^{L^t} \left| \gamma_u \right|^2 \right),
\end{equation}
where \( d_m \) denotes the distance between the \( m \)-th Eve and the BS. The proof is given as follows. 

\begin{IEEEproof}
First, we simplify the expression of \(\mathbb{E} \left[ \text{SNR}_{\text{col}} \right]\) as follows. 
\begin{equation*} 
\begin{aligned}
\mathbb{E} \left[ \text{SNR}_{\text{col}} \right] &= \mathbb{E}\left[\frac{\sum\limits_{m=1}^{M} \left| \mathbf{h}_{m}^{e}(\mathbf{T})^{H} \mathbf{w} \right|^{2}}{\sigma^{2}}\right]\\
&= \frac{1}{\sigma^{2}}\mathbb{E}\left[\sum\limits_{m=1}^{M} \left| \mathbf{h}_{m}^{e}(\mathbf{T})^{H} \mathbf{w} \right|^{2}\right]\\
&= \frac{1}{\sigma^{2}}\sum\limits_{m=1}^{M}\mathbb{E}\left[\left| \sum_{u=1}^{L^t} \sigma_u \gamma_u \right|^2\right].
\end{aligned}
\end{equation*}
Next, according to the definition of modulus \(\left| X\right|^2 = X \cdot X^*\), where \(X\) and * represents a complex number and conjugate operation. The expected expression of \(\mathbb{E}\left[\left| \sum_{u=1}^{L^t} \sigma_u \gamma_u \right|^2\right]\) can be given by
\begin{equation}\label{eq:expection proof}
\begin{aligned}
\mathbb{E}\left[\left| \sum_{u=1}^{L^t} \sigma_u \gamma_u \right|^2\right] &= \mathbb{E}\left[\left(\sum_{u=1}^{L^t} \sigma_u \gamma_u\right)\left(\sum_{v=1}^{L^t} \sigma_v \gamma_v\right)^*  \right]\\
&= \mathbb{E}\left[\sum_{u=1}^{L^t} \sum_{v=1}^{L^t}\sigma_u \gamma_u \sigma_v^* \gamma_v^* \right]\\
&= E'_{u=v} + E'_{u \neq v}.
\end{aligned}
\end{equation}
where \(E'_{u=v} = \mathbb{E}\left[\sum\limits_{u=v=1}^{L^t} \sigma_u \gamma_u  \sigma_v^* \gamma_v^*\right]\) and \(E'_{u \neq v} = \mathbb{E}\left[\sum\limits_{u \neq v}^{L^t} \sigma_u \gamma_u  \sigma_v^* \gamma_v^*\right]\). Due to \(\sigma_u, \sigma_v \sim \mathcal{CN}(0,\frac{g_0 d_m^{-\alpha}}L)\) and their mutual independence, the second term \(E'_{u \neq v} = 0\). Therefore, the first term \(E'_{u=v}\) can be expressed as 
\begin{equation*} 
\begin{aligned}
E'_{u=v} = \mathbb{E}\left[ \left| \sigma_u \right|^2 \right] \cdot \sum\limits_{u=1}^{L^t} \left| \gamma_u \right|^2 
= \frac{g_0 d_m^{-\alpha}}{L^t}\left( \sum\limits_{u=1}^{L^t} \left| \gamma_u\right|^2 \right).
\end{aligned}
\end{equation*}
Finally, By substituting \(E'_{u=v}\) and \(E'_{u \neq v}\) into Eq. \eqref{eq:expection proof}, the expression for \(\mathbb{E} \left[ \text{SNR}_{\text{col}} \right]\) can be represented as
\begin{equation*}
\mathbb{E}[\text{SNR}_{\text{col}}] = \sum_{m=1}^{M} \frac{g_0 d_m^{-\alpha}}{L^t \sigma^2} \left( \sum_{u=1}^{L^t} \left| \gamma_u \right|^2 \right).
\end{equation*}

\end{IEEEproof}

To sum up, the the expression of the objective function in Eq. \eqref{objecive init} can be rewritten as Eq. \eqref{optimal after}.
\begin{figure*}
\begin{align}
\mathbb{E}[\Delta \rm SNR_{dif}]'=\frac{g_0 d^{-\alpha}}{L^t \sigma^2} \sum_{z=1}^{M} \left| \sum_{u=1}^{L^t} \alpha_u(r_z) \gamma_u \right|^2 - \sum_{m=1}^{M} \frac{g_0 d_m^{-\alpha}}{L^t \sigma^2} \left( \sum_{u=1}^{L^t} \left| \gamma_u \right|^2 \right).\label{optimal after}
\end{align}
\hrule 
\end{figure*}
Thus, the optimization problem in Eq. \eqref{eq:optimal inital} can be simplified into:
\begin{subequations}\label{eq:simplified optimal inital1}
\begin{align}
\text{Objective:}~~&  \min_{d, \mathbf{R}} \mathbb{E}[\Delta \text{SNR}_\text{dif}]'\label{objective}\\
\text{s.t. } & \eqref{theparentequation b}, \eqref{theparentequation c}, \text{and} \eqref{theparentequation e}.
\end{align}
\end{subequations}

\section{Optimization Algorithm Design for the Security Equivalent Model of VMA}\label{sec:strategies}
In this section, we are dedicated to solving the optimization problem in Eq. \eqref{eq:simplified optimal inital1}. Due to the tight coupling features between the distance \(d\) and antenna position \(\mathbf{R}\) of the MA array at the virtual Eve, we tackle it via alternating optimization (AO). In fact, AO is a
widely applicable and empirically efficient approach for handling
optimization problems involving coupled optimization variables. It has been successfully applied to several secure
wireless communication design problems, such as maximizing the achievable secrecy rate by joint design of beamforming and
MA positions \cite{arXiv2024feng-MA-PLS}.
Based on the principles of AO, in order to obtain the optimal solution of the optimization problem, parameters (\(d\), \(\mathbf{R}\)) can be alternately optimized while keeping other variables fixed. This yields a stationary point solution of problem Eq. 
\eqref{eq:simplified optimal inital1}, which is detailed in the following two subsections.

\subsection{Distance Optimization for A Fixed \( \mathbf{R} \)}\label{op_d}
First, we find the optimal position \(d\) by fixing the antenna positions of the MA array at the virtual Eve, namely \(\mathbf{R}\). 
The distance optimization problem is formulated as:
\begin{subequations}\label{eq:optimal d}
\begin{align}
\text{Objective:}~~& \min_{d}\mathbb{E}[\Delta \text{SNR}_\text{dif}]' \label{eq:d optimal objective}\\
\text{s.t. } & \eqref{theparentequation e}.
\end{align}
\end{subequations}

When the virtual Eve moves closer to the BS, the received SNR gets larger and the objective function \eqref{eq:d optimal objective} becomes positive. Conversely, as it moves farther from the BS, the objective function \eqref{eq:d optimal objective} potentially becomes negative.
Therefore, there exists a range of values for the equivalent distance \(d\) to satisfy constraint condition \eqref{theparentequation e}. Next, solve constraint condition \eqref{theparentequation e} to obtain the maximum range of \(d\) values.
\begin{equation}
\begin{aligned}
d \leq &d_{\max}\\
&=\left[ \frac{g_0 \left( \left| \sum\limits_{u=1}^{L^t} \alpha_u(r_1) \gamma_u \right|^2 + \ldots + \left| \sum\limits_{u=1}^{L^t} \alpha_u(r_M) \gamma_u \right|^2 \right)}{\mathbb{E}[\text{SNR}_{\text{col}}] L^t \sigma^2} \right]^{\frac{1}{\alpha}}
 \label{eq:d_up}
 \end{aligned}
\end{equation}
Therefore, the constraint \eqref{theparentequation e} can be transformed into \(0 \leq d \leq d_{\max}\). In particular, above constraint is utilized to ensure the non-negativity of the objective function \eqref{eq:d optimal objective}. The objective function is convex and is a power function with respect to \(d\), which is a decreasing function of \(d\). Thus, when \(d = d_{\max}\), the objective function reaches the minimum value, and \(d_{\max}\) is the optimal solution of the  subproblem \eqref{eq:d optimal objective}. 

\subsection{Antenna Position Optimization for A Fixed \( d \)}\label{op_R}
Next, we explore the design of $\textbf{R}$ at the virtual Eve by fixing $d$, the subproblem of optimizing $\textbf{R}$ is given as follows:
\begin{subequations}\label{eq:antena opt fixed d}
\begin{align}
\text{Objective:}~~& \min_{\mathbf{R}}\mathbb{E}[\Delta \text{SNR}_\text{dif}]'\\
\text{s.t. } & \eqref{theparentequation b}, \eqref{theparentequation c}, \eqref{theparentequation e}.
\end{align}
\end{subequations}
Since the objective function with respect to the optimization variable \( \mathbf{R} \) is non-convex and difficult to solve, we introduce a slack variable \(v\) to transform the objective function into a convex function. As a result, let
\begin{equation}
 v_m \geq \left| \sum_{u=1}^{L^t} \alpha_u(r_m) \gamma_u \right|^2, \quad m = 1, \ldots, M .
 \label{equ:v_m}
\end{equation}
However, the above expression contains a modulus squared operator, which also makes the expression complex and difficult to solve. Next, we handle the modulus squared operation by letting \( \sum\limits_{u=1}^{L^t} \alpha_u(r_m) \gamma_u = p_m + jq_m \), where \( p_m \) and \( q_m \) represent the real part and imaginary part, respectively. Then we can get 
\begin{equation}
p_m = R_{\rm op} \left[ \sum\limits_{u=1}^{L^t} \alpha_u(r_m) \gamma_u \right],
\label{equ:p_m}
\end{equation}
and 
\begin{equation}
q_m = I_{\rm op} \left[ \sum\limits_{u=1}^{L^t} \alpha_u(r_m) \gamma_u \right],
\label{equ:q_m}
\end{equation}
where $R_{\rm op}$ and $I_{\rm op}$ denote the operations of taking the real part and the imaginary part, respectively. 
Thus, Eq. \eqref{equ:v_m} can be rewritten as:
\begin{equation}
v_m \geq \left| \sum_{u=1}^{L^t} \alpha_u(r_m) \gamma_u \right|^2 =  p_m^2 + q_m^2 .
\label{equ: p_m^2 + q_m^2}
\end{equation}
Moreover, the constraint \eqref{theparentequation c} is non-convex. Since the MA array at the virtual Eve, modeling the different positions of $M$ Eves with single FPA, is a uniform linear array, we only need to consider that the two adjacent antennas satisfy the minimum distance constraint in sequence. Accordingly, the constraint \eqref{theparentequation c} can be rewritten as:
\begin{equation}
\mathbf{r}_n - \mathbf{r}_{n-1} \geq D_{\min} \quad n = 2, \dots ,N.\label{theparentequation c'}
\end{equation}
To sum up, the optimization problem \eqref{eq:antena opt fixed d} can be re-expressed as
\begin{subequations}\label{eq:simplified optimal inital}
\begin{align}
\text{Objective:}~~& \min_{\mathbf{R},v_m} \frac{g_0 d^{-\alpha}}{L^t \sigma^2} \sum_{m=1}^{M} v_m  - \mathbb{E}[\text{SNR}_{\text{col}}]\label{cvx}\\
\text{s.t. } & \frac{g_0 d^{-\alpha}}{L^t \sigma^2} \sum_{m=1}^{M} v_m  - \mathbb{E}[\text{SNR}_{\text{col}}] \geq 0 \label{theparentequation v>0},\\
& \eqref{theparentequation b}, \eqref{equ:p_m}, \eqref{equ:q_m},\eqref{equ: p_m^2 + q_m^2}, \eqref{theparentequation c'}. 
\end{align} 
\end{subequations}
It can be noticed that the objective function \eqref{cvx} with respect to the optimization variable \( v_m\) becomes a convex function. Furthermore, constraints \eqref{theparentequation b}, \eqref{equ:p_m}, \eqref{equ:q_m}, \eqref{equ: p_m^2 + q_m^2}, \eqref{theparentequation c'} and \eqref{theparentequation v>0} are convex, which can be easily solved by the CVX toolbox. 

\subsection{Overall algorithm}\label{Overall algorithm}
In this subsection, we propose a AO algorithm to solving the complex non-convex problem \eqref{eq:simplified optimal inital1}. The pseudo-code of optimizing the antenna positions of the MAs at the virtual Eve, i.e.,$\textbf{R}$, and equivalent distance between the BS and the virtual Eve, i.e., $d$, by AO is described as follows. 

\begin{algorithm}[!htp]
\label{alg:AO}
    \caption{{Joint Optimization of Equivalent Distance and Antenna Positions of MAs based on AO (\textbf{JO\_EDAP\_AO})}}
    \label{alg:AO}
    \renewcommand{\algorithmicrequire}{\textbf{Initialize:}}
    \renewcommand{\algorithmicensure}{\textbf{Output:}}
    \begin{algorithmic}[1]
        \REQUIRE the maximum iteration number $I_{\rm ter}$, previous position of VMA $\mathbf{R}$, Expectation distance $d$
         \STATE $d \gets$ outputted through Eq. \eqref{eq:d_up} with $\mathbf{R}$
         \STATE $\mathbf{R} \gets$ outputted by using CVX to solve problem \eqref{cvx} 
       \ENSURE $d, \mathbf{R}$ until the maximum iteration number is reached  
    \end{algorithmic}
\end{algorithm}

\textbf{Sensitivity analysis:} In optimization problem \eqref{eq:simplified optimal inital1}, the sensitivity of the objective function to the optimization variables \( d \) and \( \mathbf{R} \) differs significantly. Specifically, the objective function is more sensitive to changes in \( d \), while it is relatively less sensitive to changes in \( \mathbf{R} \). This sensitivity difference can be attributed to their distinct roles in the objective function. In particular, \( d \), being part of the power function, directly influences the value of the objective function in an exponential manner. Therefore, even small changes in \( d \) can significantly impact the outcome of the objective function. In contrast, \( \mathbf{R} \) appears in the phase of a complex exponential function. After taking the modulus, changes in \( \mathbf{R} \) have a relatively smaller effect on the actual value of the objective function.

\textbf{Complexity Analysis:} Given the non-convexity of the problem, the problem \eqref{eq:simplified optimal inital1} is divided into two subproblems and can be solved by AO algorithm. First, in section \ref{op_R}, CVX toolbox is utilized by solving the subproblem of \eqref{eq:antena opt fixed d} based on interior-point methods. Combining with the results in \cite{TCCN2021sun-complexity}, the computational complexity is \(O(N^{3}\log(\epsilon^{-1}))\) in each iteration, where \(N\) is the number of variables, and \(\epsilon\) is the iteration accuracy. Next, in section \ref{op_d}, we can obtain a expression of \(d\), given in Eq. \eqref{eq:d_up}, with the help of the closed-form solution for \(d\), resulting in a complexity of \(O(1)\). Thus the total computational complexity of AO algorithm is \(O(I_{\text{iter}}N^{3.5}\log(\epsilon^{-1}))\), where $I_{\rm iter}$ denotes the maximum number of iterations.

\section{Evaluations}\label{sec:evaluations}
In this section, numerical results are provided through MATLAB simulator to verify the effectiveness and accuracy of the proposed equivalent model of the secrecy rate between $M$ collusion Eves and an Eve with $M$ MAs, namely   deviation minimization. In particular, on the one hand, we investigate the impact of the Eves' size, path-loss exponent, noise power, antenna moving range at the virtual Eve, and antennas' size at the virtual Eve on the accuracy approximation of the proposed equivalent model, respectively. On the other hand, the single optimizations for the antenna positions and number of the MAs at the virtual Eve, and the distance between the BS and the virtual Eve are regarded as the baselines for the performance comparison. The BS transmits signals in 28GHz, then the wavelength is 0.0107m. In addition, the results are the average of 5000 ones.

\subsection{Simulation Settings}
In the scenario, the BS utilizes a planar array with 8 MAs to transmit confidential information to the Bob, in presence of 4 randomly distributed Eves colluding to intercept the transmitted information. The MAs is a linear array and number of MAs at the virtual Eve is 4. The structure of MA arrays at the BS and virtual Eve is given in Fig. \ref{fig:sketch_2}. The minimum distance between any two MAs is set as \(0.5 \lambda\), and the moving range of each MA is \(4 \lambda\). The distance between the BS and Bob is 20m. Also, the distance between the BS and Eves follows a Gaussian distribution with a mean of 40m and a variance of 5m. The BS's transmission power is 10 mW, and the noise power is 0.5 mW. Other parameters are set according to Table \ref{tab:simulation_parameters}.

\begin{table}[h!]
\centering
\caption{\small Simulation Parameters and Values}\label{table:parameter}
\begin{tabular}{lp{5cm}l} 
\toprule
Parameter & Description & Value \\
\midrule
\( \lambda \) & wavelength & 0.0107m \\
\( N \) & number of MAs at the BS & $8^{\cite{WCL2024mei-MA}}$\\
\( M \) & number of the Eves & 4 \\
\( d_{\rm bob-bs} \) & distance between the Bob and the BS & 20m \\
\( P_t \) & transmit power & 10mW \\
\( \sigma^2 \) & noise power & 0.5mW \\
\( L \) & number of transmitting/receiving paths & 4 \\
\( g_0 \) & average channel gain at reference of 1m & 30dB \\
\( \alpha \) & path-loss exponent & 4 \\
\( A \) & movement range of each MA & \( 4\lambda \) \\
\( D_{\text{min}} \) & minimum distance between antennas & \( {0.5\lambda}^{\cite{CL2024hu-MA}} \) \\
\( I_{\rm ter} \) & maximum number of iterations & 25 \\
\bottomrule
\end{tabular}
\label{tab:simulation_parameters}
\end{table}

\subsection{Algorithm Convergence Analysis}
\begin{figure}[htbp]
  \centering 
  \includegraphics[width=0.35\textwidth]{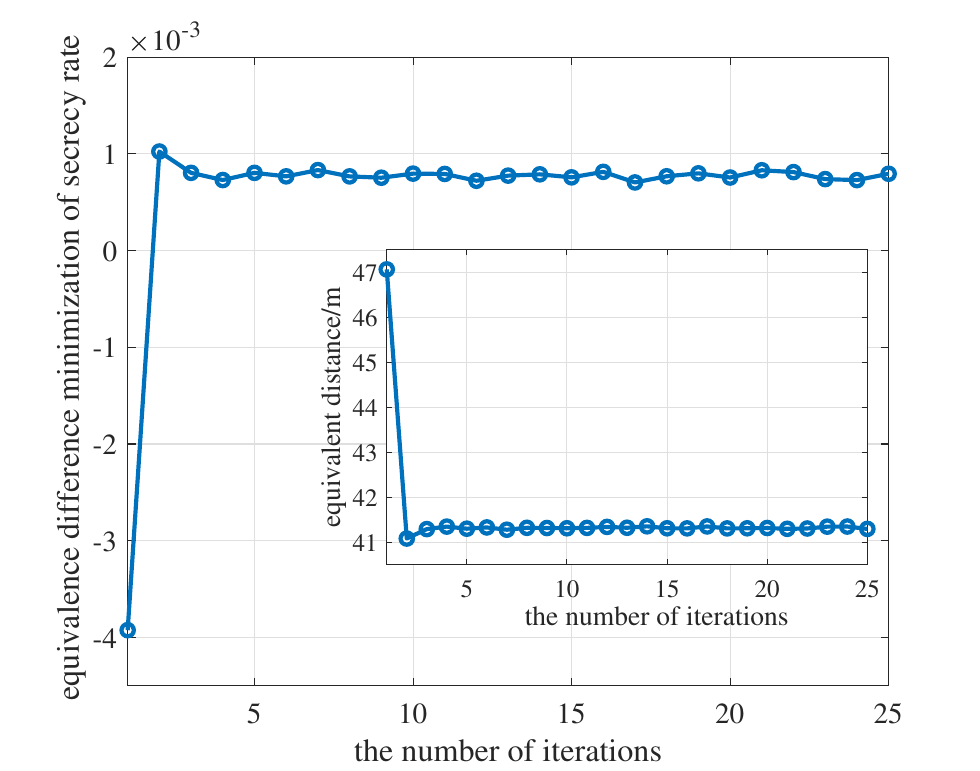}
  \caption{\small expected convergence curve} 
  \label{fig:E_d_iter}
\end{figure}

With the settings of parameters in Table 
\ref{tab:simulation_parameters}, in Fig. \ref{fig:E_d_iter}, based on CVX tools, we first demonstrate the convergence of the proposed algorithm for optimizing antenna positions of the MAs at the virtual Eve. In particular, after several iterations, the objective function converges rapidly to a stable value. 

\begin{figure}[htbp]
\centering  
\subfigure[\small original antenna position]{
\includegraphics[width=0.35\textwidth]{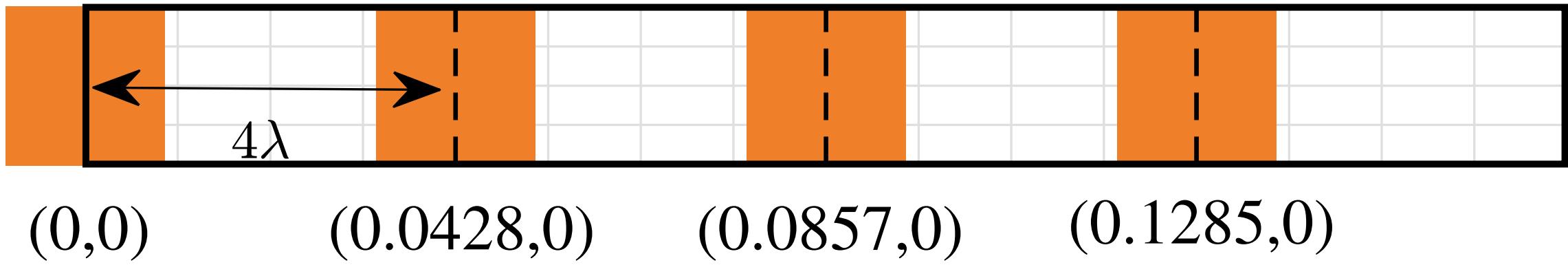}\hspace{5mm}\label{sub:antenna pre}}\\
\subfigure[\small optimal antenna position]{
\includegraphics[width=0.35\textwidth]{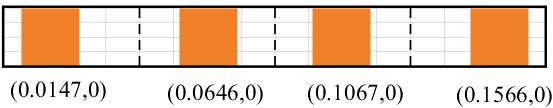}\label{sub:antenna op}}
\caption{\small antenna positions of MAs before and after moving}
\label{fig:Position comparison of Antenna array} 
\end{figure}

From the Fig. \ref{fig:E_d_iter}, we can find that the objective function is negative in the first iteration. The reason behind this phenomenon is based on sensitivity analysis in \ref{Overall algorithm}, which can be described from following two aspects: 1) In the first iteration, the initial \( d \) is calculated by using the initial coordinates of MAs' positions. As can be seen from the subgraph in Fig. \ref{fig:E_d_iter}, the value of \( d \) at this point is relatively large, which means that the virtual Eve is far from the BS, resulting in the objective function being negative. 2) given the value of \( d \), the antenna position matrix \( \mathbf{R} \) can be optimized, which yields the optimal positions of MAs \( \mathbf{R}^* \) at this stage. Since \( \mathbf{R} \) has a smaller influence on the objective function, even when using \( \mathbf{R}^* \) computes the objective function value, it cannot offset the impact of \( d \) and causes a negative feature of the objective function. After executing several iterations, \( d \) converges to a stable value, and the optimal antenna position \( \mathbf{R} \) can be obtained, which ultimately converges to a positive value. 
Compared with the initial antenna positions in Fig. \ref{sub:antenna pre}, it can be observed that the positions of the 4 MAs get a significant improvement for reaching their optimal positions and obtaining the best equivalent effect. Based on above achieved the optimal antenna positions of MAs at the virtual Eve, we can derive directly the equivalent distance between the BS and the virtual Eve via Eq. \eqref{eq:d_up} while satisfying constraint condition. To sum up, the optimization problem (\ref{eq:simplified optimal inital}) can be effectively solved.

\subsection{Performance Analyses of Equivalent Distance and Equivalence Difference Minimization of Secrecy Rate}
\subsubsection{The Impact of the Number of Eves on the Equivalent Distance and Equivalence Difference Minimization of Secrecy Rate}
\begin{figure}[htbp]
\centering  
\subfigure[\small equivalent distance \emph{vs.} the number of Eves]{
\includegraphics[width=0.35\textwidth]{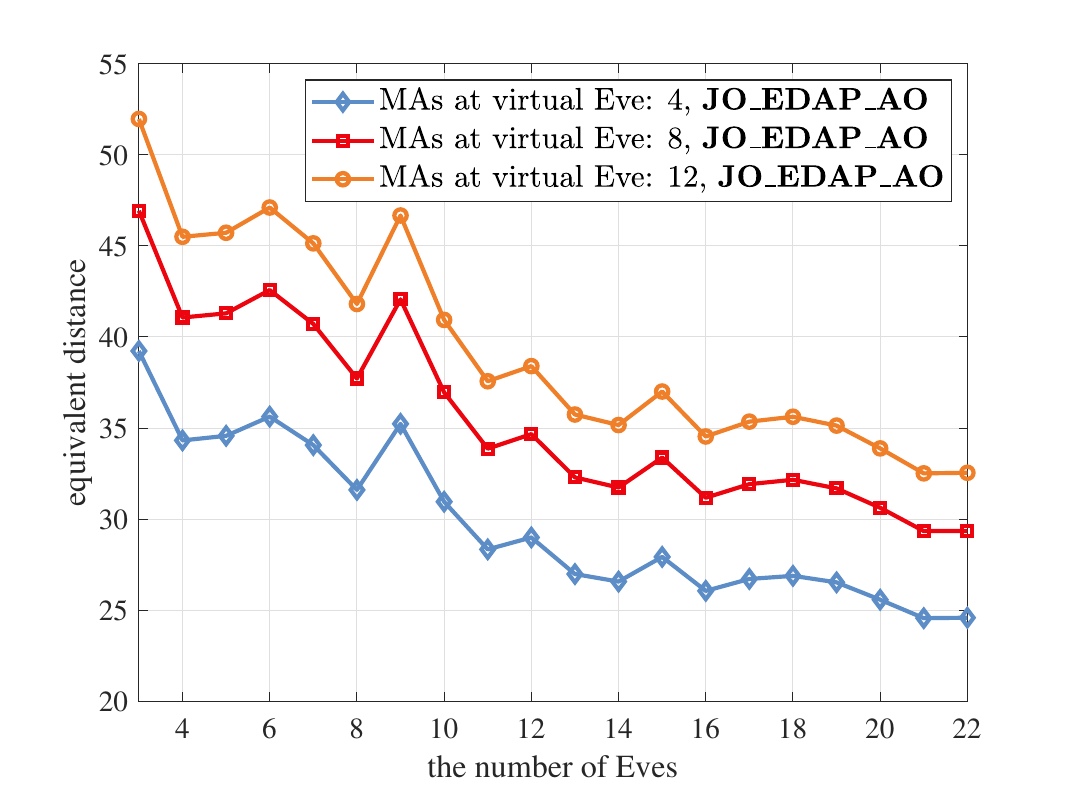}\label{sub:d_evenum}}\\
\subfigure[\small equivalence difference minimization of secrecy rate \emph{vs.} the number of Eves]{
\includegraphics[width=0.35\textwidth]{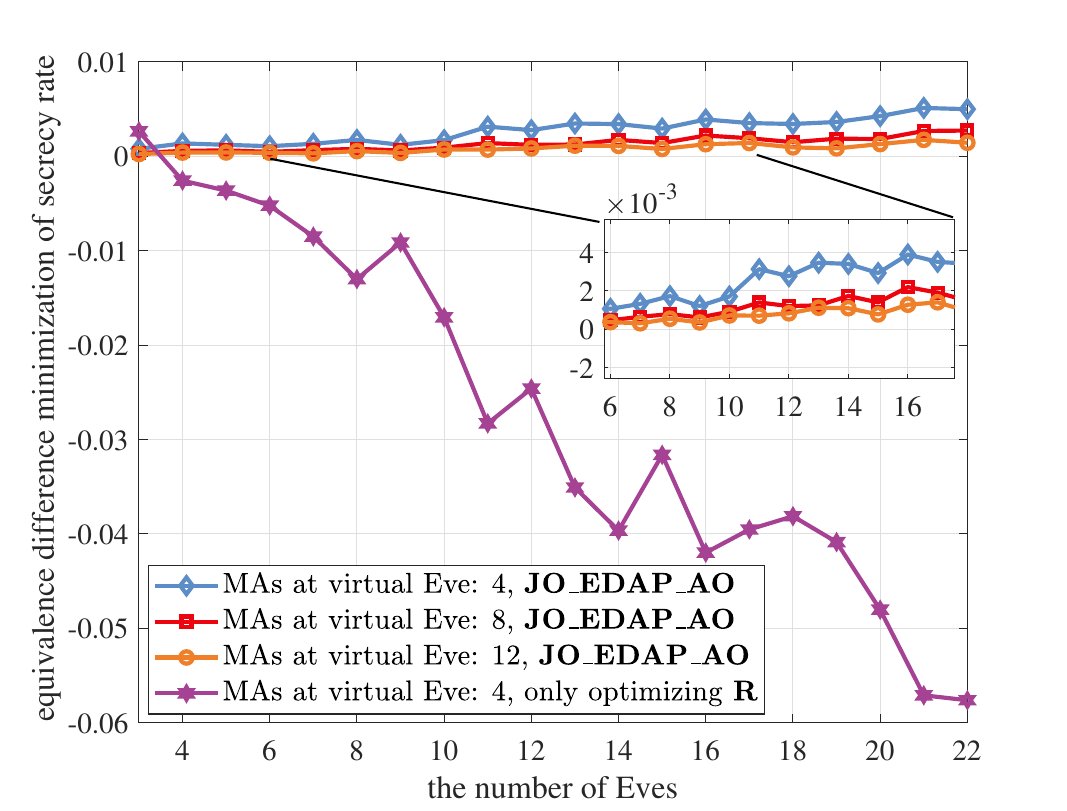}\label{sub:E_evenum}}
\caption{\small relationship between equivalent effect and the number of Eves}
\label{fig:E_d_evenum} 
\end{figure}

The distance $d$ between the BS and the virtual Eve is an crucial factor for minimizing the gap of secrecy rate achieved by $M$ Eves with single FPAs and an virtual Eve with $M$ MAs in the optimization problem \eqref{eq:optimal inital}. In Fig. \ref{sub:d_evenum}, we validate the impact of the number of Eves on the equivalent distance $d$, under different settings of MAs' size at the virtual Eve. It can be noticed that, on the one hand, the equivalent distance shows a downward trend over the number of Eves increasing, since a larger number of Eves means that there are more Eves colluding with each other for intercepting the confidential information, which leads to a higher channel capacity based on the expression of Eq. \eqref{eq:channel capacity eves}, and a smaller equivalent distance is needed for obtaining a stronger strength of eavesdropping signal at the virtual Eve. And, the equivalent distance gets larger when the virtual Eve has more MAs for modeling the positions of the Eves. The reason behind this phenomenon is that more MAs at the virtual Eve can compensate the strength of eavesdropping signal caused by the equivalent distance, by adding the spatial DoF via antenna movement. 

On the other hand, as shown in Fig. \ref{sub:E_evenum}, the equivalence difference minimization of secrecy rate obtained by $M$ Eves with single FPA and a virtual Eve with $M$ MAs gets worse, when the number of Eves increases. In particular, with the settings of 4 MAs, 8 MAs and 12 MAs at the virtual Eve, the equivalence difference minimization of secrecy rate decreases. In addition, 12 MAs at the virtual Eve can achieve the optimal the equivalence difference minimization of secrecy rate, i.e., more closest to 0. The corresponding reason is that more antennas exist at the virtual Eve, more spatial DoF can be improved for modeling the different locations of the Eves by moving more MAs. Moreover, as one of baseline, with the settings of 4 MAs, the distance between the BS and the virtual Eve is averaged, and the decline in equivalent performance becomes more pronounced as the number of Eves increases, the equivalence difference minimization of secrecy rate has the worst performance. Due to the increment in the number of Eves, \(\text{SNR}_{\text{col}}\) continues to rise, while using the average distance cannot effectively increase \(\text{SNR}_{\text{veve}}\), resulting in a degradation of the equivalence performance.

\subsubsection{The Impact of the path-loss exponent and noise power on the Equivalent Distance and Equivalence Difference Minimization of Secrecy Rate}

\begin{figure}[htbp]
\centering  
\subfigure[equivalence difference minimization of secrecy rate \emph{vs.} path-loss exponent]{
\includegraphics[width=0.38\textwidth]{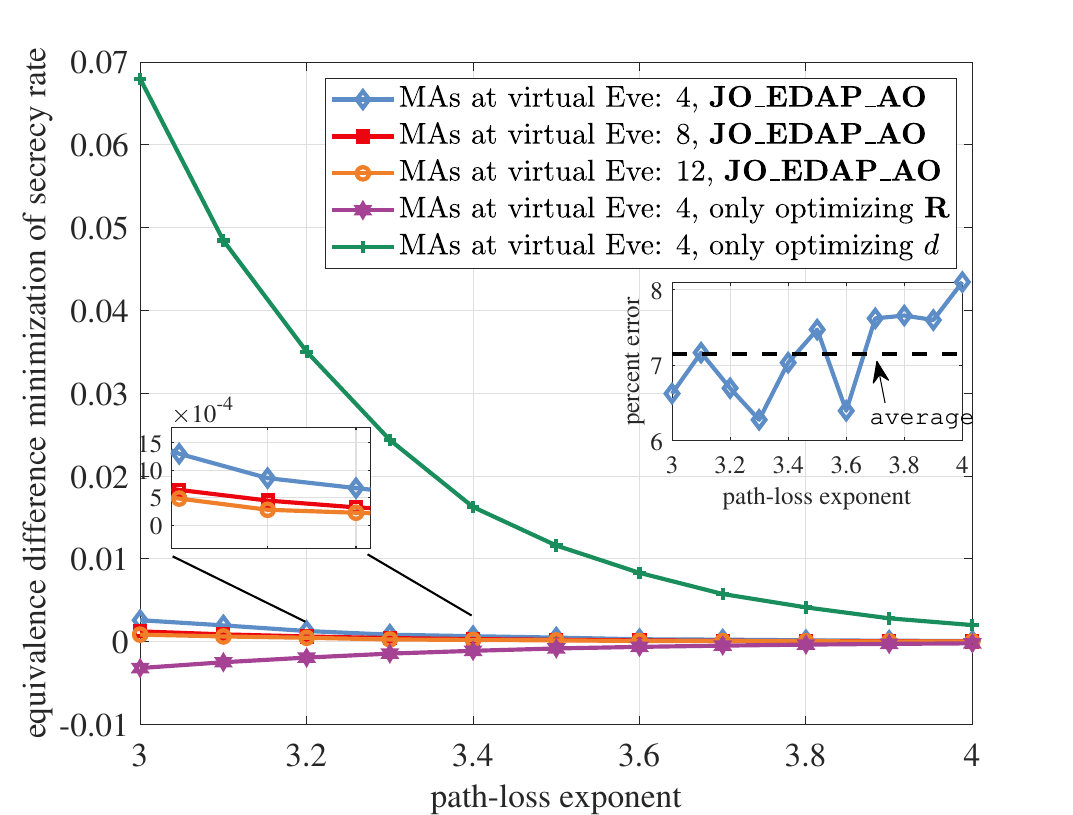}\label{sub:E_a}}\\
\subfigure[equivalence difference minimization of secrecy rate \emph{vs.} noise power]{
\includegraphics[width=0.38\textwidth]{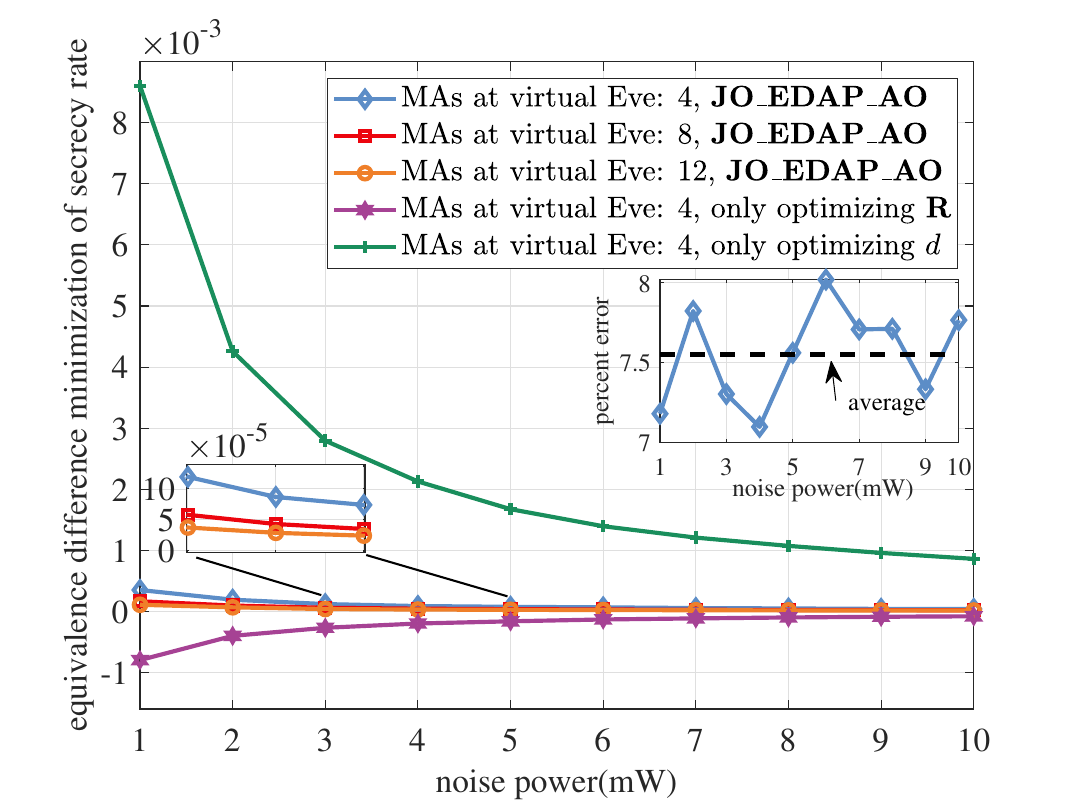}\label{sub:E_noisepower}}
\caption{relationship between equivalence difference minimization of secrecy rate and path-loss exponent, noise power}
\label{fig:E_a_noisepower} 
\end{figure}

Fig. \ref{sub:E_a} and Fig. \ref{sub:E_noisepower} validate the impact of different path-loss exponent and noise power on the equivalence difference minimization of secrecy rate, under different setting of MAs' size at the virtual Eve, respectively. 
On the one hand, as shown in Fig. \ref{sub:E_a}, the equivalence difference of secrecy rate before and after establishing equivalent model decreases as the increase of path-loss exponent. One of possible reasons is due to the decrement in the expected magnitude of SNR before and after establishing equivalent model of approximating secrecy rate in Subsection \ref{sub:equivalence difference minimization}, leading to a drop in values of the equivalence difference. The subplots in the figure can further prove the findings, which describe the percentage error achieved by \(\left(\mathbb{E}[\text{SNR}_{\text{col}}] - \mathbb{E}[\text{SNR}_{\text{veve}}]\right) /\mathbb{E}[\text{SNR}_{\text{col}}]\), and show no clear trend with variable changes. In particular, corresponding percentage error is about \(7\%\) on the average. In addition, the green line represents the results by only optimizing the equivalent distance \(d\) and using the initial antenna positions of the MAs. Compared with the results achieved by optimizing both equivalent distance $d$ and antenna positions of MAs at the virtual Eve $\textbf{R}$, the former results in a significant error on equivalent effect, which further demonstrates the necessity of MAs' position optimization. The purple line represents the results by only optimizing the antenna positions of the MAs with the average distance of \(M\) Eves. The equivalence difference of the secrecy rate before and after establishing equivalent model in Subsection \ref{sub:equivalence difference minimization} is less than zero, which means the SNR for colluding Eves with single FPAs is lower than that of an virtual Eve with MAs, since the distance from the BS is too far, resulting in a very low SNR. Finally, other three lines (i.e., 4 MAs, 8 MAs, and 12 MAs.) indicate that the more antennas of the MAs there are, the better the equivalence difference minimization of the secrecy rate before and after establishing equivalent model in Subsection \ref{sub:equivalence difference minimization}.
Similarly, as shown in Fig. \ref{sub:E_noisepower}, over the noise power increasing, similar conclusions were also achieved. 

\begin{figure}[htbp]
\centering  
\subfigure[\small equivalent distance \emph{vs.} path-loss exponent]{
\includegraphics[width=0.35\textwidth]{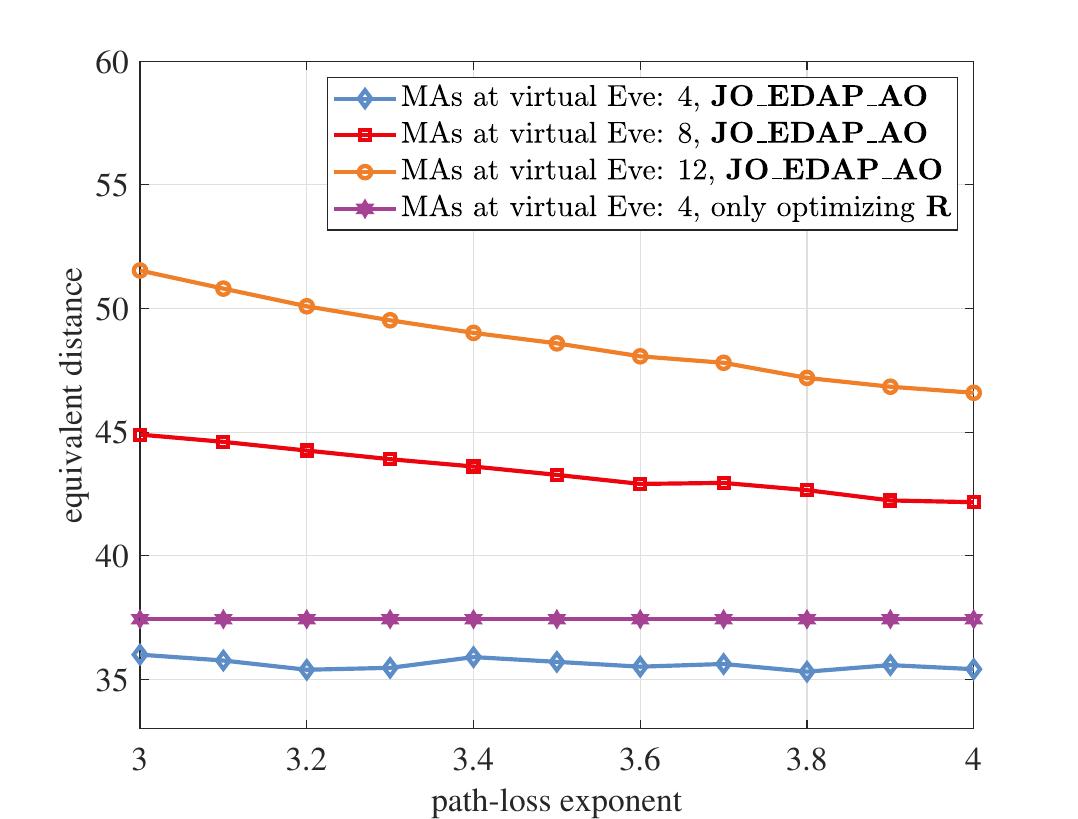}\label{sub:d_a}}
\subfigure[\small equivalent distance \emph{vs.} noise power]{
\includegraphics[width=0.35\textwidth]{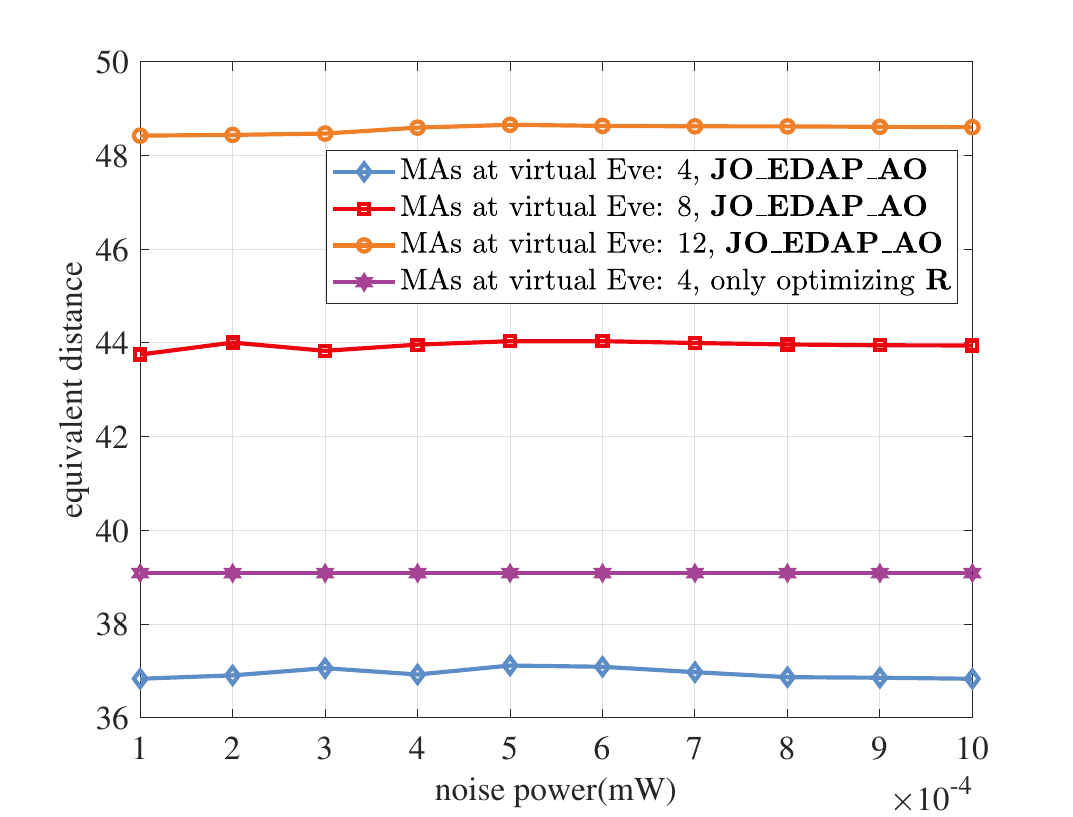}\label{sub:d_noisepower}}
\caption{\small relationship between equivalent distance and path-loss exponent, noise power}
\label{fig:d_a_noisepower} 
\end{figure}

Considering the impact of different path-loss exponents on the The distance $d$ between the BS and the virtual Eve, known as an crucial factor for minimizing the gap of secrecy rate achieved by $M$ Eves with single FPAs and an virtual Eve with $M$ MAs in the optimization problem \eqref{eq:optimal inital} of the optimization problem \eqref{eq:optimal inital}, in Fig. \ref{sub:d_a}, we can find that the equivalent distance decreases over the path-loss increasing, since the virtual Eve can compensate for the signal attenuation caused by a larger path-loss exponent by shortening the distance between the BS and the virtual Eve. In particular, the equivalent distance gets larger as the number of MAs at the virtual Eve increases. The corresponding reason behind this phenomenon is that, given a fixed path-loss exponent, if the virtual Eve has more antennas, more channel characteristics between the BS and the virtual Eve can be modeled by adjusting antenna positions of MAs at the virtual Eve rather than the equivalent distance. Therefore, a larger equivalent distance between the BS and the virtual Eve can be allowed to exist.

Furthermore, the impact of different settings of noise power on the equivalent distance is shown in Fig. \ref{sub:d_noisepower}. It can be noticed that the equivalent distance has a nearly constant for all settings of noise powers, which proves that the noise power is not a main factor of affecting the equivalent distance. Moreover, a larger value of the noise power also gets a longer equivalent distance, corresponding reasons are similar to these of Fig. \ref{sub:d_a}.

\subsubsection{The Impact of MAs' Moving Range on Equivalent Distance and Equivalence Difference Minimization of Secrecy Rate}
\begin{figure}[htbp]
\centering  
\subfigure[\small equivalent distance \emph{vs.} moving range of MAs]{
\includegraphics[width=0.35\textwidth]{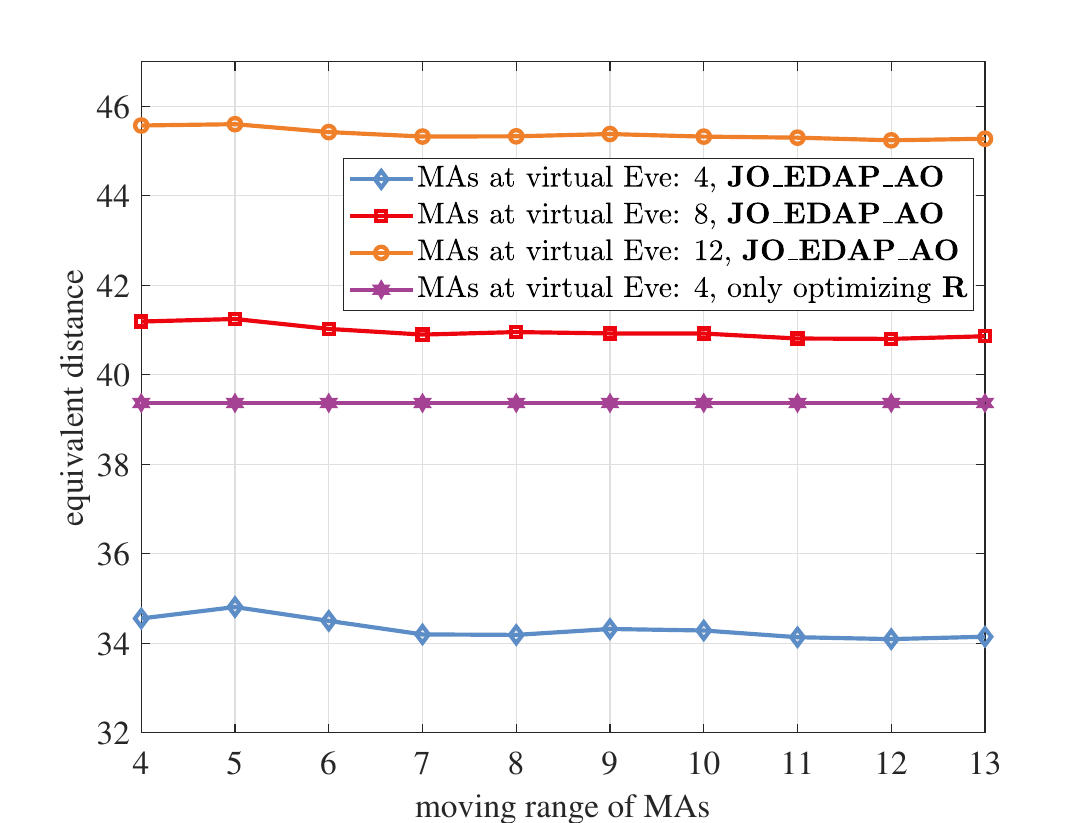}\label{sub:d_range}}
\subfigure[\small equivalence difference minimization of secrecy rate \emph{vs.}  moving range of MAs]{
\includegraphics[width=0.35\textwidth]{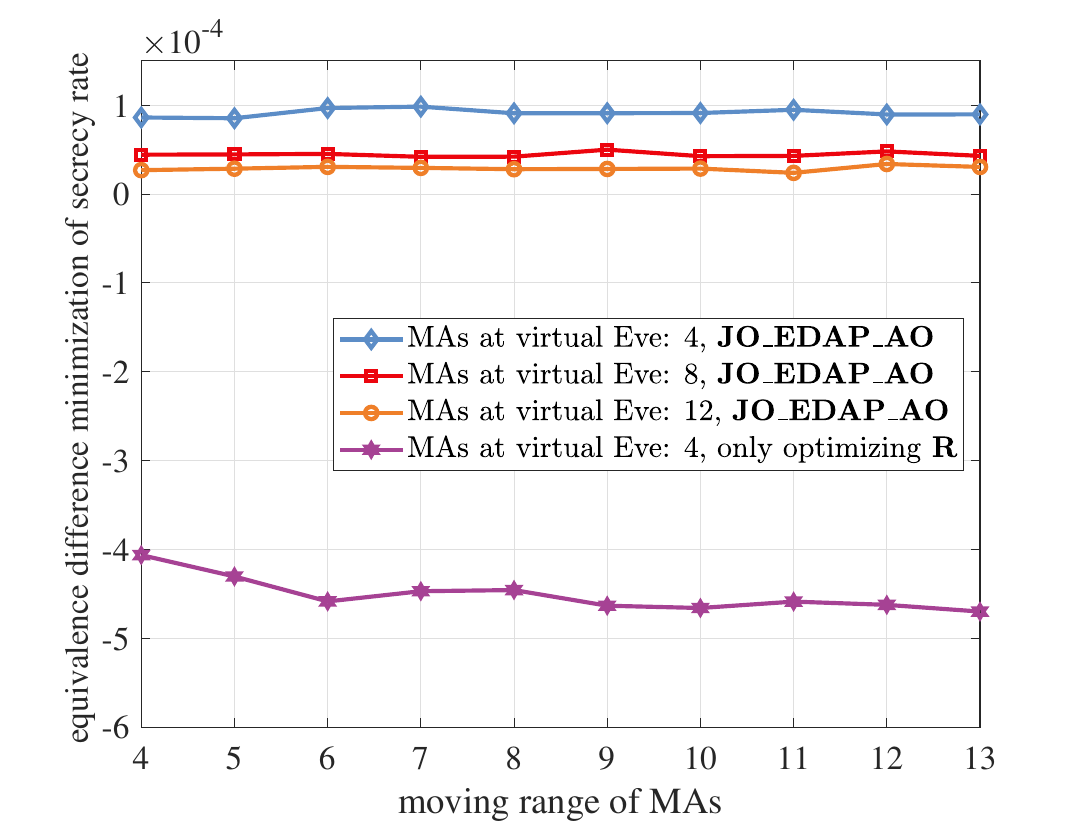}\label{sub:E_range}}
\caption{\small relationship between equivalent distance/equivalence difference minimization of secrecy rate and moving range of MAs}
\label{fig:E_d_range} 
\end{figure}

As shown in Fig. \ref{fig:E_d_evenum} to Fig. \ref{fig:d_a_noisepower}, it can be noticed that the number of MAs at the virtual Eve has a great influence on the equivalent distance and equivalence difference minimization of secrecy rate. Accordingly, in this Subsection, we first observe how the equivalent distance and equivalence difference minimization of secrecy rate vary with different settings of moving range of MAs, respectively, in Fig. \ref{sub:E_range} and Fig. \ref{sub:d_range}. Similar to the results in Fig. \ref{sub:d_noisepower}, it can be observed that the moving range of MAs at the virtual Eve has almost no effects on above two aspects. One possible reason is that although a larger moving range can model more channel characteristics between the BS and the virtual Eve by adjusting antenna positions of MAs, the optimal/approximate optimal channel gain of each antenna can be achieved by only moving a smaller distance. In particular, the moving distance of all MAs is on the average, which also proves that the moving range of MAs is not a major factor of affecting the equivalent distance and equivalence difference minimization of secrecy rate.

\begin{figure}[htbp]
  \centering 
  \includegraphics[width=0.52\textwidth]{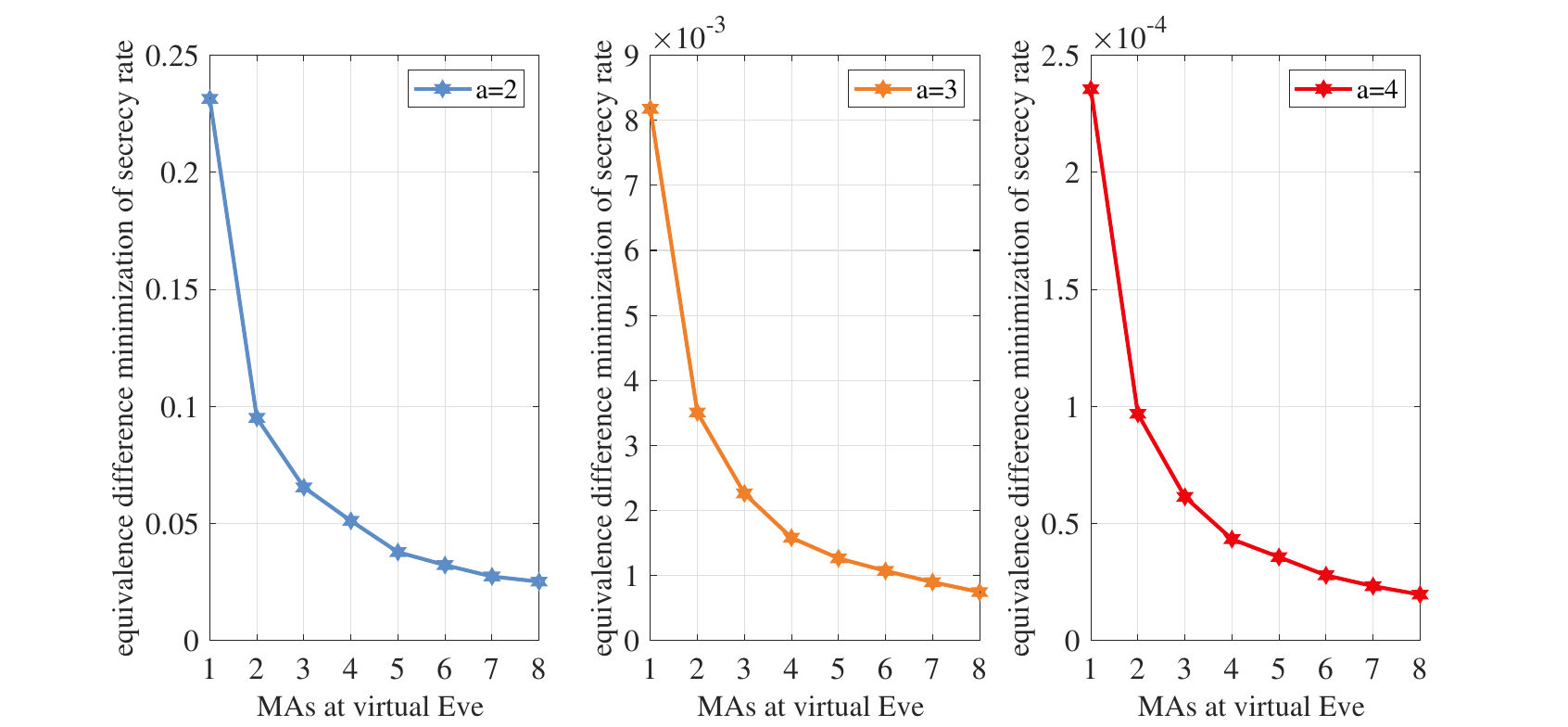}
  \caption{\small the impact of the number of MAs at the virtual Eve on equivalence difference minimization of secrecy rate} 
  \label{fig:E_antennanum}
\end{figure}

Next, considering three typical environments with path-loss exponent being 2, 3, and 4, namely free space, suburban areas with some obstacles, and urban areas with many obstacles, 
in Fig. \ref{fig:E_antennanum}, we observe how the equivalence difference minimization of secrecy rate varies with the number of MAs at the virtual Eve, under different settings of path-loss exponent. It can be noticed that from Fig. \ref{fig:E_antennanum} increasing the number of MAs at the virtual Eve can improve the equivalence performance, namely minimizing the gap between the secrecy rates obtained by $M$ Eves with single FPA and a virtual Eve with $M$ MAs. This is because that the increment of the number of MAs leads to a larger distance between the BS and virtual Eve, which improves the equivalent performance indirectly. In addition, a larger path-loss exponent can further achieve the equivalence difference minimization of secrecy rate, the reasons are similar to these of Fig. \ref{sub:E_a}.

In particular, the equivalence difference decreases significantly with 2 MAs, and gradually converges when the number of MAs is greater than 7. From the perspective of energy consumption caused by moving antennas and equivalence difference minimization of secrecy rate, it is essential to make a trade-off between them by determining/optimal the number of MAs needing to move\footnote{In this paper, we only validate the impact of MAs' size on the equivalence difference minimization of secrecy rate. Optimization of the number of MAs will be  analyzed in future work.}.

\begin{figure}[htbp]
  \centering 
  \includegraphics[width=0.35\textwidth]{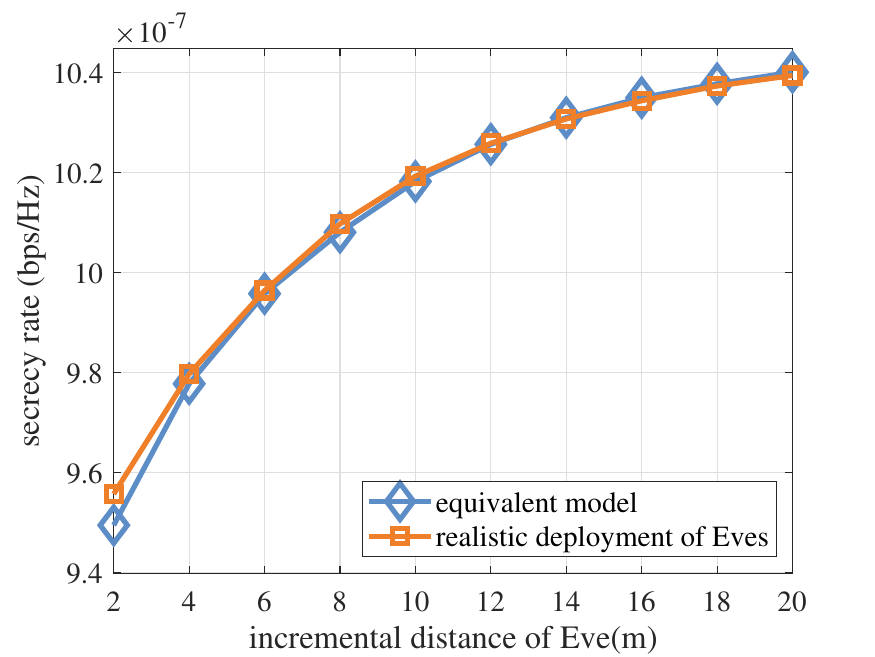}
  \caption{\small the secrecy rate of two models} 
  \label{fig:sec-evedis}
\end{figure}

\begin{table}[htbp]
\centering
\caption{\small Secrecy rate comparison of equivalent model vs. realistic deployment of Eves}
\label{tab:secrecy rate comparison of equivalent model vs. realistic deployment of Eves}
\setlength{\tabcolsep}{1.1mm}{
\begin{tabular}{*{6}{c}}
  \toprule
  \multirow{2}*{Eve position} & \multicolumn{2}{c}{secrecy rate(bps/Hz)} & \multirow{2}*{error} \\  
  \cmidrule(lr){2-3}
  & Eves with single FPAs & an Eve with MAs & \\
  \midrule
  \([50\quad50.5\quad53]\) & $9.557 \times 10^{-7}$ & $9.495 \times 10^{-7}$ & $0.65\% $ \\
  \([58\quad58.5\quad61]\) & 1.019 $\times 10^{-6}$ & $1.018 \times 10^{-6}$ & $0.1\%$ \\
  \([68\quad68.5\quad71]\) & $1.039 \times 10^{-6}$ & $1.040 \times 10^{-6}$ & $-0.06\%$ \\
  \bottomrule
\end{tabular}}
\end{table}

Finally, we observe the difference of secrecy rates, which are achieved by $M$ Eves with single FPAs and a virtual Eve with $M$ MAs, respectively.
In particular, the number of Eves is 3, and the distance between the BS and the Bob is 30m. Initially, the distances between them and the BS are set to 48m, 48.5m, and 51m. By gradually increasing the distance between the BS and all Eves but without changing the relative position of all Eves, on the one hand, We can find that the secrecy rate obtained by the virtual Eve with $M$ MAs is very close to that of $M$ Eves with single FPAs. From Table \ref{tab:secrecy rate comparison of equivalent model vs. realistic deployment of Eves}, it can also be noticed that the error before and after equivalence is minimal, less than 1\%. The secrecy rate in the simulation is quite low, which may be attributed to the broadcast nature of the wireless channel and the lack of BF design for the users. In particular, the difference gets smaller over the distance between Eves and BS increasing, and the difference is minimal on the average. On the other hand, the secrecy rate increases when the distance between the BS and all Eves gets larger, since the Bob has more better channel quality than the Eves.
To sum up, the equivalent model of the secrecy rate between $M$ collusion Eves and a virtual Eve with $M$ MAs in Subsection \ref{subsec:equ model} is effective.

\section{Conclusions and Future Works}\label{sec:conclusions}
In this paper, to derive a closed-form expression for the secrecy rate in the presence of multiple randomly distributed eavesdroppers (Eves), we established an equivalent model that approximates this secrecy rate by considering a single Eve equipped with a movable antenna (MA) array. Furthermore, to minimize the deviation between the two types of secrecy rates, we formulated an optimization problem. To address the non-convexity of this problem, we employed an alternative optimization (AO) approach to jointly design the equivalent distance between the transmitter and the virtual Eve, as well as the antenna positions of the MAs at the virtual Eve. Simulation results verified the effectiveness and accuracy of the proposed equivalence model.
In practice, obtaining perfect channel state information (CSI) of various moving devices is challenging, which affects the performance of physical layer security (PLS) methods. Therefore, in future work, we will focus on designing MA-based PLS methods. Moreover, integrated sensing and communication (ISAC) technologies bring new performance enhancements to wireless communications; thus, we will also focus on ISAC and MA-based PLS in our future research.

\bibliographystyle{IEEEtran}
\bibliography{references}

\end{document}